\PassOptionsToPackage{table}{xcolor}

\documentclass[11pt]{article}
\usepackage{jheppub}
\usepackage{graphicx} 
\usepackage{amsmath}
\usepackage{amsfonts}
\usepackage{hyperref}
\usepackage{braket}
\usepackage{physics}
\usepackage{comment}
\usepackage{dsfont}
\usepackage{hhline}
\usepackage{float}
\usepackage{bm}
\usepackage{tikz}
\usepackage{enumitem}
\usepackage{mathtools}
\usepackage{mathrsfs}
\usetikzlibrary{shapes,arrows,cd,chains,decorations.markings,decorations.pathmorphing,calc,positioning,patterns}

\usepackage{xcolor}
\usepackage{booktabs, array, makecell, multirow}
\usepackage{adjustbox, subcaption}

\tikzset{
->-/.style args={#1rotate#2}{decoration={markings, mark=at position #1 with {\arrow[scale=1.5,rotate = #2 ]{stealth}}}, postaction={decorate}}
}
\tikzset{
-r-/.style args={#1rotate#2}{decoration={markings, mark=at position #1 with {\arrow[scale=1,rotate = #2 ]{>}}}, postaction={decorate}}
}

\DeclareMathOperator{\bbR}{\mathbb{R}}
\DeclareMathOperator{\bbC}{\mathbb{C}}

\DeclareMathOperator{\calC}{\mathcal{C}}
\DeclareMathOperator{\calD}{\mathcal{D}}

\DeclareMathOperator{\calH}{\mathcal{H}}

\DeclareMathOperator{\calQ}{\mathcal{Q}}
\DeclareMathOperator{\calR}{\mathcal{R}}
\DeclareMathOperator{\calO}{\mathcal{O}}
\DeclareMathOperator{\calS}{\mathcal{S}}
\DeclareMathOperator{\calK}{\mathcal{K}}
\DeclareMathOperator{\calL}{\mathcal{L}}

\DeclareMathOperator{\calV}{\mathcal{V}}

\DeclareMathOperator{\Ker}{\mathrm{Ker}}

\DeclareMathOperator{\HT}{{\mathrm{HT}}}
\DeclareMathOperator{\TFD}{{\mathrm{TFD}}}

\DeclareFontFamily{U}{mathx}{\hyphenchar\font45}
\DeclareFontShape{U}{mathx}{m}{n}{<-> mathx10}{}
\DeclareSymbolFont{mathx}{U}{mathx}{m}{n}
\DeclareMathAccent{\widebar}{0}{mathx}{"73}

\newcommand{\floor}[1]{\left\lfloor #1 \right\rfloor }

\newmuskip\pFqskip
\mathchardef\pFcomma=\mathcode`, 

\title{Observables in Schr\"odinger CFTs:\\How Aliens Built the Pyramids}

\author[1]{Mathieu Boisvert,}
\author[1]{Shehab Hossam Fadda,}
\author[1,2]{Justin Kulp,} 
\author[2,3]{and Ramtin M.Yazdi}

\affiliation[1]{Yang Institute for Theoretical Physics, Stony Brook University, Stony Brook, NY 11794, USA}
\affiliation[2]{Simons Center for Geometry and Physics, Stony Brook University, Stony Brook, NY 11794, USA}
\affiliation[3]{Department of Physics and Astronomy, Stony Brook University, Stony Brook, New York 11794-3800, USA}
\emailAdd{mathieu.boisvert@stonybrook.edu}
\emailAdd{shihab.fadda@stonybrook.edu}
\emailAdd{jkulp@scgp.stonybrook.edu}
\emailAdd{ramtin.mohasselyazdi@stonybrook.edu}

\abstract{We discuss the algebraic structure of observables in Schr\"odinger CFTs. These operators have zero mass (or particle number) and generically transform in staggered ``pyramid representations'' built from ``alien operators,'' as we explain with the doubled state-operator correspondence. We comment on implications for the space of non-relativistic CFTs, thermal physics, and generalize the exceptional symmetry conservation laws of Bekaert, Meunier, and Moroz, and Golkar and Son. We show that alien operators are analogous to double-twist operators in Lorentzian CFT, with systematic cross-channel corrections from massless particles when they exist.}

\begin{document}

\maketitle

\section{Introduction}
Schr\"odinger CFTs describe non-relativistic fixed points with dynamical critical exponent $z=2$ \cite{NR0Mass, Henkel:1993sg, Henkel:2003pu, Duval:2024eod, Baiguera:2023fus}. They are of both experimental and theoretical relevance, with applications to: fermions at unitarity, cold atoms in a harmonic trap, and the BCS-BEC crossover \cite{hornreich1975critical, grinstein1981anisotropic, roberts1998resonant,regal2004observation, nussinov2004bcs, Son:2005rv, nussinov2006triviality, Mehen:2007dn, Nishida:2007pj, chang2007unitary, von2007bec, Gegenwart:2008ttt, Nishida:2010tm}; nuclear physics EFTs \cite{Kaplan:1998tg, Kaplan:1998we, Mehen:1999nd, Kaplan:2009kr, Kobach:2018nmt, Arav:2019tqm}; Langevin dynamics \cite{Ardonne:2003wa, Son:2013rqa, Hoyos:2013qna, Chapman:2015wha, Arav:2016akx, Yan:2022dqk, DearQuantizers}; non-relativistic limits of relativistic QFTs \cite{Beg:1984yh,Jackiw:1991je,Bergman:1991hf,Chapman:2020vtn}; Lorentz-breaking deformations of relativistic CFTs \cite{Cardy:1992tq, Antunes:2026jjf}; null-reductions of Lorentzian CFTs \cite{Duval:1994qye, Duval:2008jg, NullDefects}; large quantum number expansions \cite{nikolic2007renormalization, Bekaert:2011qd, Kravec:2018qnu, Favrod:2018xov, Kravec:2019djc, Hellerman:2023myh, Lee:2026azy, Beane:2026mxi}; and string theory and holography \cite{Son:2008ye, Goldberger:2008vg, Balasubramanian:2008dm, Barbon:2008bg, Maldacena:2008wh, Kachru:2008yh, Herzog:2008wg, Gomis:2020fui, Gomis:2020izd, Maxfield:2022hkd, Dorey:2022cfn, Dorey:2023jfw, Mouland:2023gcp}. 

By definition, Schr\"odinger CFTs in ($d+1$)-dimensions are invariant under the Schr\"odinger group
\begin{equation}\label{eq:SchroGroup}
    S_d := (SL(2,\bbR) \times SO(d)) \ltimes H_{d}\,.
\end{equation}
The $SL(2,\bbR)$ is generated by time translation $P_0$, a $z=2$ dilatation $D$, and a timelike special conformal transformation $C_0$, the $SO(d)$ acts by spatial rotations $M_{ij}$, and the $H_d$ is a Heisenberg group generated by spatial translations $P_i$, boosts $K_i$, and a central mass element $M$. We refer to Appendix \ref{app:algebra} for commutation relations and \cite{NR0Mass} for more details.

Despite their prevalence in a diverse range of physical scenarios, formal universal treatments of Schr\"odinger CFTs have gone largely undeveloped (but see \cite{Nishida:2007pj, Goldberger:2008vg, Goldberger:2014hca} for important foundational works), precluding an axiomatic ``bootstrap'' approach to Schr\"odinger CFTs. 

One issue of particular importance concerns the definition and properties of ``normal ordered'' operators, i.e. operators of the schematic form $\calO^\dagger_a\!\calO_b$.\footnote{The well-definedness of such normal ordered operators in general strongly coupled Schr\"odinger CFTs follows from a (complete) state-operator correspondence \cite{NR0Mass} and associated non-renormalization theorems \cite{Bergman:1991hf, Chapman:2020vtn, NR0Mass}.} These operators annihilate the ``harmonic trap'' vacuum state $\Omega$ from the left and the right, and include all densities for actual physical Hermitian observables, like the number density $n(x) = (\phi^\dagger\phi)(x)$, probability current $j_i(x)$, and stress tensor $T_{ij}(x)$. Every global symmetry current is also of this type.

These normal ordered operators should play an important role in a bootstrap program for Schr\"odinger CFTs. In particular, in \cite{NR0Mass} it was argued that they play an analogous role in the Schr\"odinger OPE to double-twist operators in relativistic CFTs. When there is only one $M=0$ state in the harmonic trap spectrum (the vacuum state $\Omega$), their scaling dimensions are (in a sense) classical, and when additional $M=0$ states are in the harmonic trap spectrum, they obtain anomalous dimensions similar to the $1/\ell$ corrections to double-twist scaling dimensions from non-trivial exchanges in the $t$-channel OPE \cite{Simmons-Duffin:2016wlq}. They also (non-coincidentally) have connections to light-ray operators, and are directly related in null-reductions.

In \cite{Bekaert:2011qd, Golkar:2014mwa}, Bekaert, Meunier, Moroz, Golkar, and Son noted some additional intriguing properties: Schr\"odinger CFTs possess an infinite number of normal ordered $M=0$ operators which are neither primaries nor descendants, of the schematic form $\calO^\dagger \! 
\overset{\leftrightarrow}{\partial}_{\mu}^{\raisebox{-1.2ex}{$\scriptstyle n$}}  \! \calO$, dubbed ``alien operators'' in \cite{Golkar:2014mwa}. It was observed that alien operators and families of primaries combine into ``pyramid'' representations, with links between conservation laws at different levels in the pyramid. In this paper, we provide a complete treatment of these pyramid representations, showing that they form reducible but indecomposable (staggered) modules for the Schr\"odinger algebra, and showing that the entire sector of $M=0$ observables generically decomposes into pyramids.

\vskip 0.5cm
\noindent This is a technical note on algebraic and representation theoretic aspects of the neutral sector, necessary for understanding conformal blocks and other $M=0$ data. The outline and summary of the remainder of the paper is as follows:
\begin{enumerate}\setlength\itemsep{0em}
    \item[{\hyperref[sec:TFDStates]{$\S$.2.}}] We start in Section \ref{sec:TFDStates} by reviewing the relevant representation theory of Schr\"odinger CFTs and the thermofield double construction. In Section \ref{sec:LWRs}, we enumerate the physical/admissible representations of the Schr\"odinger algebra $S_d$, using techniques analogous to usual radial quantization. Admissible reps include massive and massless modules, obtained as (quotients of) Verma modules:
    \vskip -0.75cm
    \begin{alignat}{3}
        \calL_{\Delta,m} &:= \calV_{\Delta,m} 
        \qquad &&\Delta > d/2\,, \quad m > 0\,,\\
        \calL_{\Delta,0} &:= \calV_{\Delta,0}/(P_i \sim 0)
        \qquad &&\Delta > 0\,.
    \end{alignat}
    \vskip -0.25cm
    In Section \ref{sec:TensorProduct}, we discuss the thermofield double state-operator correspondence and non-renormalization theorems, justifying the subsequent algebraic manipulations. Finally, in Section \ref{sec:MassNotZero}, we combine these two sections to understand generic composite operators (where $m_1 \neq -m_2$), giving a tensor product decomposition of massive modules:
    \vskip -0.25cm
    \begin{equation}
        \calL_{\Delta_1, m_1} \otimes \calL_{\Delta_2, -m_2} = \bigoplus_{\ell = 0}^\infty n_\ell \calL_{\Delta_1 + \Delta_2 + \ell, m_1 - m_2}\,.
    \end{equation}
    \vskip 0.0cm
    \item[{\hyperref[sec:Pyramids]{$\S$.3.}}]  We dedicate the entirety of Section \ref{sec:Pyramids} to understanding the normal ordered $M=0$ sector, pyramid representations $\calS_{\Delta}$, and their properties. We start in Section \ref{sec:Intuition} by giving a bottom-up construction of the normal ordered $M=0$ sector, generalizing the discussions of \cite{Bekaert:2011qd, Golkar:2014mwa}, and explicitly constructing the states/representations in the tensor product
    \vskip -0.5cm
    \begin{equation}\label{eq:pyramidTensor}
        \calL_{\Delta_1, m} \otimes \calL_{\Delta_2, -m} = \bigoplus_{\ell = 0}^\infty p_\ell \calS_{\Delta_1 + \Delta_2 + 2\ell}\,.
    \end{equation}
    In Section \ref{sec:rigorous} we give an intrinsic top-down definition of the pyramids $\calS_{\Delta}$, characterizing them rigorously as a direct limit of short exact sequences of massless Verma modules $\calV_{\Delta,0}$. In Section \ref{sec:conservation} we turn to applications; we discuss symmetry current conservation rules, constraints on the space of NR CFTs, and show how the pyramids give descent relations among conservation laws. In Section \ref{sec:TFD}, we briefly discuss some consequences for thermal calculations and write the finite-temperature TFD state $\ket{\ket{\TFD}}_{\beta,\mu}$, of interest for thermal or holographic calculations, in the coupled basis.
    \vskip 0.25cm
    \item[{\hyperref[sec:Bootstrap]{$\S$.4.}}] In Section \ref{sec:Bootstrap} we discuss the renormalization of normal-ordered $M=0$ operators in the presence of massless particles. We show that normal-ordered ``alien operators'' in the $s$-channel OPE play a role analogous to double-twist primaries in Lorentzian CFT, and that the anomalous dimensions receive systematic corrections from $t$-channel exchanges by resumming logarithms into cross-channel blocks.
    \vskip 0.25cm
    \item[{\hyperref[app:algebra]{\underline{$\mathscr{A}$.}}}] In Appendix \ref{app:algebra} we provide conventions for the Schr\"odinger algebra. We give a careful discussion of Lorentzian and Euclidean conventions, and explain BPZ conjugation for non-relativistic CFTs.
    \vskip 0.25cm
    \item[{\hyperref[sec:explicitPyramid]{\underline{$\mathscr{B}$.}}}] In Appendix \ref{sec:explicitPyramid} we explicitly construct all states in the tensor product decomposition \eqref{eq:pyramidTensor}. Crucially, we use this to compute Clebsch-Gordan coefficients (to our knowledge, for the first time) and get explicit formulas for the TFD states in Section \ref{sec:TFD}.
    \vskip 0.25cm
    \item[{\hyperref[sec:PyramidConstructive]{\underline{$\mathscr{C}$.}}}] In Appendix \ref{sec:PyramidConstructive} we discuss additional properties of pyramid representations $\calS_{\Delta}$. We give a third definition of pyramid representations, arguing that they are uniquely characterized by the following physical consistency conditions: a bounded below operator spectrum, maximal operator content compatible with Ward identities, and compatibility with the OPE.
\end{enumerate}

\section{States in the TFD and Normal Ordered Operators}\label{sec:TFDStates}
Essentially by definition, the Hilbert space $\calH$ of a Schr\"odinger CFT decomposes into non-negative energy (projective) unitary irreducible representations of the (centrally extended) Schr\"odinger group $S_d$, which we will call ``admissible'' irreps for short. i.e., 
\begin{equation}\label{eq:Hilb}
    \calH = \bigoplus_{\Delta,\rho,m} N_{\Delta,\rho,m} \calL_{\Delta,\rho,m}\,.
\end{equation}
The non-negative energy condition is defined by the harmonic trap Hamiltonian
\begin{equation}\label{eq:HTHamiltonian}
    H_{\HT} := P_0 + \omega^2 C_0\,.
\end{equation}
Our goal in this section is to describe the admissible representations, and setup and compute the tensor product decomposition of the TFD Hilbert space into Schr\"odinger representations. The entirety of Section \ref{sec:Pyramids} is dedicated to the special subcase with $M=0$, describing operators like the number density $n(x) = (\phi^\dagger\phi)(x)$.

\subsection{Lowest Weight Representations for Schr\"odinger}\label{sec:LWRs}
By a Wick rotation and coordinate transform, it is possible to map the Schr\"odinger CFT from the harmonic trap spacetime to the Euclidean plane, such that the trap Hamiltonian $H_{\HT}$ becomes a dilatation $D$, closely mirroring standard CFT. See Appendix \ref{app:algebra} for details.

The Schr\"odinger algebra $\mathfrak{sch}_d$ admits a triangular decomposition graded by $D$
\begin{equation}
\mathfrak{sch}_d =
\begin{array}[t]{ccccccccc}
    \mathfrak{s}_{-2} & \oplus 
    & \mathfrak{s}_{-1} & \oplus 
    & \mathfrak{s}_{0} & \oplus 
    & \mathfrak{s}_{1} & \oplus 
    & \mathfrak{s}_{2} \\[1pt]
    C_0 &&&&D&&&& P_0\\[1pt]
    && K_i && M && P_i &&\\[1pt]
    &&&& M_{ij} &&&&
\end{array}\,.
\end{equation}
We define the lower and upper-triangular pieces for $D$ to be $\mathfrak{s}_- := \mathfrak{s}_{-2} \oplus \mathfrak{s}_{-1}$ and $\mathfrak{s}_+ := \mathfrak{s}_{1} \oplus \mathfrak{s}_2$ respectively \cite{Dobrev:2013kha}. 

The admissible representations appearing in $\calH$ turn out to all be primary lowest weight modules built from a primary lowest weight vector $\ket{\Delta,\rho, m}$. By definition, a lowest weight vector diagonalizes $\mathfrak{s}_0$
\begin{equation}
    D \ket{\Delta,\rho,m} = \Delta \ket{\Delta,\rho,m}
    \quad \text{and} \quad
    M \ket{\Delta,\rho,m} = m \ket{\Delta,\rho,m}\,,
\end{equation}
transforms in the $\rho$-representation of $SO(d)$, and is annihilated by all of $\mathfrak{s}_-$
\begin{equation}
    C_0 \ket{\Delta,\rho,m} = 0 
    \quad \text{and}\quad
    K_i \ket{\Delta,\rho, m} = 0\,,
\end{equation}
see \cite{NR0Mass, Perroud:1977qh, dobrev1997lowest, Dobrev:2013kha}. The full Verma module of descendants is generated by acting freely with the raising operators $P_0$ and $P_i$ in $\mathfrak{s}_+$. E.g., a scalar Verma module is of the form:\footnote{Since the rotations $M_{ij}$ never appear on the RHS of any commutators in the Schr\"odinger algebra (except other rotations), we restrict to scalar lowest-weights without essentially any interesting change of structure.\label{footnote:noSpin}}
\begin{equation}\label{eq:GeneralVerma}
    \calV_{\Delta,m} := \bbC\{P_0^{n_0} P_1^{n_1} \cdots P_d^{n_d}\ket{\Delta,m} \,|\, n_\mu \geq 0\}\,.
\end{equation}
The Verma modules have an obvious ``level''-grading, corresponding to the $D$-weight of a vector
\begin{equation}
    (\calV_{\Delta,m})_k := \bbC\{P_0^{n_0} P_1^{n_1} \cdots P_d^{n_d}\ket{\Delta,m} \,|\, 2n_0 + n_1 + \dots + n_d = k \}\,.
\end{equation}

\begin{figure}[t]
\centering
    \begin{minipage}[b]{0.48\textwidth}
    \centering
    \begin{tikzcd}[cramped, column sep=-1.0em, row sep = scriptsize]
    	&&&& {\ket{\Delta,m}} \\
    	&&& {P_i \ket{\Delta,m}} \\
    	&& {P_i^2\ket{\Delta,m}} && {P_{0}\ket{\Delta,m}} \\
    	& {P_i^3 \ket{\Delta,m}} && {P_0P_i \ket{\Delta,m}} \\
    	{P_i^4 \ket{\Delta,m}} & {} & {P_0 P_i^2 \ket{\Delta,m}} & {} & {P_0^2 \ket{\Delta,m}}
    	\arrow[from=1-5, to=2-4]
    	\arrow[from=1-5, to=3-5]
    	\arrow[from=2-4, to=3-3]
    	\arrow[from=2-4, to=4-4]
    	\arrow[from=3-3, to=4-2]
    	\arrow[from=3-3, to=5-3]
    	\arrow[from=3-5, to=4-4]
    	\arrow[from=3-5, to=5-5]
    	\arrow[from=4-2, to=5-1]
    	\arrow[no head, from=4-2, to=5-2]
    	\arrow[from=4-4, to=5-3]
    	\arrow[no head, from=4-4, to=5-4]
    \end{tikzcd}
    \subcaption*{\newline\large$\hphantom{\ket{\Delta,m}} \mathcal{L}_{\Delta,m}\,,\quad \Delta >d/2\,.$}
    \end{minipage}\hfill
    \begin{minipage}[b]{0.48\textwidth}
    \centering
\begin{tikzcd}[cramped, column sep=1.0em, row sep = scriptsize]
	& {\ket{\Delta,m}} \\
	{\vphantom{P_0\ket{\Delta}}} \\
	& {P_{0}\ket{\Delta,m}} \\
	{\vphantom{P_0\ket{\Delta}}} \\
	& {P_0^2 \ket{\Delta,m}}
	\arrow["{P_i \sim 0}"'{pos=0.4}, color={rgb,255:red,214;green,92;blue,92}, from=1-2, to=2-1]
	\arrow[from=1-2, to=3-2]
	\arrow["{P_i \sim 0}"'{pos=0.4}, color={rgb,255:red,214;green,92;blue,92}, from=3-2, to=4-1]
	\arrow[from=3-2, to=5-2]
\end{tikzcd}
\subcaption*{\newline\large$\hphantom{\ket{\Delta,m}} \mathcal{L}_{\Delta,0}\,,\quad \Delta > 0\,.$}
    \end{minipage}
    \caption{Left, the module $\calL_{\Delta,m} \cong \calV_{\Delta,m}$ with $\Delta > d/2$ describing ``genuine massive operators.'' It has a ``half-pyramid'' shape. A null-state appears at level $2$ at the unitarity bound $\Delta=d/2$. Right, the module $\calL_{\Delta,0}$ describing ``genuine massless operators.'' It effectively behaves as an $SL(2,\bbR)$ module. All $P_i$ (red) lead to states which are quotiented out of $\calV_{\Delta,0}$, leaving states/operators spatially homogeneous/independent.}
    \label{fig:easyReps}
\end{figure}
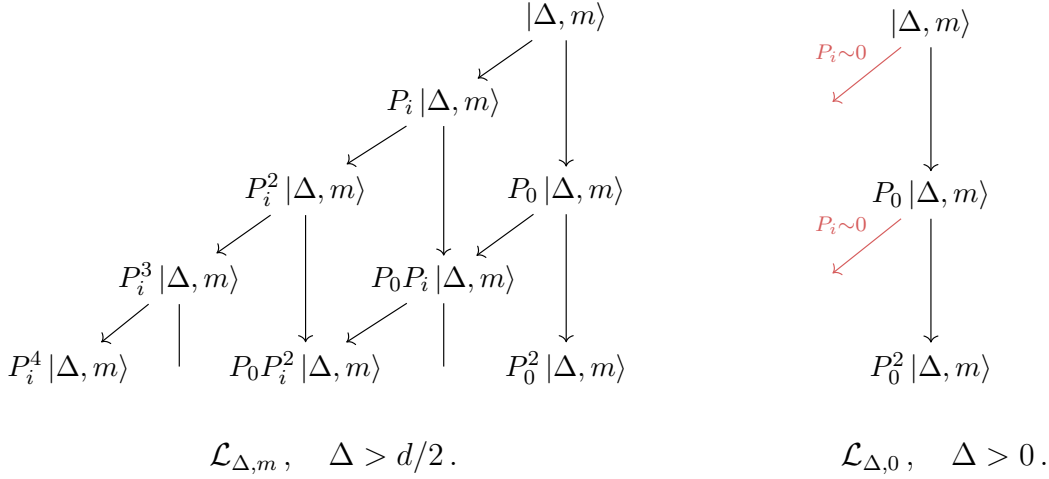

Unitarity (of real-time generators) and irreducibility place additional constraints on the representations $\calV_{\Delta,m}$. Pulled back to the Euclidean plane, Hermitian conjugation corresponds to the standard Shapovalov-like inner product on the Schr\"odinger algebra, acting on generators by:
\begin{equation}
    D^\dagger = D\,,\quad
    P_0^\dagger = C_0\,,\quad
    P_i^\dagger = K_i\,,\quad
    M_{ij}^\dagger = - M_{ij}\,,\quad
    M^\dagger = M\,.
\end{equation}
Thus, in many cases, we find singular vectors which must be quotiented out to obtain proper unitary irreducible representations. The results are as follows:
\begin{enumerate}\setlength\itemsep{0em}
    \item \textbf{If $\bm{m > 0}$:} 
    \begin{enumerate}\setlength\itemsep{0em}
        \item \textbf{If $\bm{\Delta > d/2}$,} then $\calV_{\Delta,m}$ is an irreducible module
        \begin{equation}
            \calL_{\Delta,m} 
                := \calV_{\Delta,m}
                \quad \text{$m>0$\, and\, $\Delta > d/2$}\,.
        \end{equation}
        \item \textbf{If $\bm{\Delta = d/2}$,} then there is a singular vector $v := (P_0 - \vec{P}^2/2m)\ket{\Delta,m}$ at level $2$, which can be quotiented out to give the irreducible module
        \begin{equation}
            \calL_{\frac{d}{2},m} := \calV_{\frac{d}{2},m}/(v \sim 0)\,.
        \end{equation}
        $\Delta = d/2$ is called the massive unitarity bound \cite{Nishida:2007pj, Goldberger:2014hca}.
    \end{enumerate} 
    \item \textbf{If $\bm{m = 0}$:} 
    \begin{enumerate}\setlength\itemsep{0em}
        \item \textbf{If $\bm{\Delta > 0}$,} then all $v_i := P_i \ket{\Delta,m}$ are singular vectors at level $1$ and can be quotiented out to give the irreducible module
        \begin{equation}
        \calL_{\Delta,0} := \calV_{\Delta,0}/(v_i \sim 0)\,.
    \end{equation}
        These correspond to ``genuine massless operators'' in \cite{NR0Mass}.
        \item \textbf{If $\bm{\Delta = 0}$,} then all descendants are singular, and we quotient to the trivial representation $\calL_{0,0}$. $\Delta = 0$ is called the massless unitarity bound \cite{Pal:2018idc, NR0Mass}.
    \end{enumerate}
    \item \textbf{If $\bm{m < 0}$:} there are no admissible representations.
\end{enumerate}
We draw these reps in Figure \ref{fig:easyReps}. No additional unitarity constraints come from considering spinning modules $\rho \neq \mathds{1}$. Normal ordered operators, like symmetry currents or the number density $n(x)$, do not appear in this list because they do not correspond to states in the physical harmonic trap Hilbert space $\calH$.\footnote{There are also discrete families of disconnected representations below the unitarity bounds when $m \geq 0$, see Theorem 1 of \cite{Dobrev:2013kha}. By the analyticity considerations in Section 3.3.1 of \cite{NR0Mass}, and the construction of group representations in \cite{Perroud:1977qh}, we do not expect them to be physically relevant in Schr\"odinger CFTs.}

\subsection{Tensor Product Hilbert Space and Composites}\label{sec:TensorProduct}
The state-operator correspondence for Schr\"odinger CFTs identifies local operators with states in the TFD Hilbert space $\calH \otimes \calH^*$, i.e. endomorphisms of $\calH$ after an identification \cite{NR0Mass}. The rest of this section, and Section \ref{sec:Pyramids}, is dedicated to understanding the transformation properties of these tensor products under the Schr\"odinger group. By \eqref{eq:Hilb}, this essentially reduces to understanding the tensor product of the lowest weight representations $\calL_{\Delta_1,\rho_1,m_1}$ appearing in $\calH$ with the reps $\calL_{\Delta_2,\rho_2,m_2}^*$ appearing in $\calH^*$.

Our primary interest in understanding the structure of the TFD Hilbert space comes from a desire to understand the composite ``normal ordered'' local operators of Schr\"odinger CFTs. In \cite{NR0Mass}, it was shown that the non-relativistic OPE defines an, essentially canonical, normal ordered composite operator
\begin{equation}
    \calO_{1}^\dagger\!\calO_2
        \quad\text{with}\quad \Delta = \Delta_{1} + \Delta_2\,,
\end{equation}
dual to the state $\ket{1}\otimes \ket{2^*}$. A key feature for this definition is that the OPE between pure creation and pure annihilation operators is completely regular if non-trivial $M=0$ states are absent from the HT spectrum $\calH$, and thus there are no anomalous dimensions. In generic examples, $M=0$ states are not believed to be present \cite{NR0Mass}. This can be used to argue for the classical addition/tensoring of scaling dimensions/quantum numbers, even in strongly interacting theories. Thus, by understanding the tensor product Hilbert space, we will understand the collection of composite normal ordered operator algebra of Schr\"odinger CFTs since these composite operators essentially behave as if the theory is free.

By definition, primary local operators $\calO_1^\dagger(0)$ and $\calO_2(0)$ transform under the adjoint Schr\"odinger action by
\begin{alignat}{2}
    [D, \calO_1^\dagger(0)] &= \Delta_1\! \calO_1^\dagger(0)\,,\quad
    [M, \calO_1^\dagger(0)] &&= m_1\! \calO_1^\dagger(0)\,,\\
    [D, \calO_2(0)] &= \Delta_2\! \calO_2(0)\,,\quad
    [M, \calO_2(0)] &&= -m_2\! \calO_2(0)\,.
\end{alignat}
$\calO_1^\dagger$ and its descendants (at the origin) transform like the lowest weight module $\calL_{\Delta_1,\rho_1,m_1}$, and $\calO_2$ and its descendants (at the origin) transform like the lowest weight module
\begin{equation}
    \calL^*_{\Delta_2,\rho_2,m_2} \cong \calL_{\Delta_2,\rho_2^*,-m_2}\,.
\end{equation}
Since $m_2>0$, the conjugate lowest weight module is obviously not unitary for the aforementioned notion of unitarity. This is perfectly sensible: we don't want/expect local annihilation operators $\calO_{2}(0)$ to create states when acting on the HT vacuum $\ket{\Omega}$.

We will not complete the exhaustive exercise listing all conceivable tensor products $\calL_{\Delta_1,\rho_1,m_1} \otimes \calL_{\Delta_2,\rho_2,m_2}^*$ appearing in $\calH \otimes \calH^*$ in this paper. Instead, we specialize to the two most important cases, and focus on (1+1)d where the structures are already visible.\footnote{There is essentially no interesting information lost by not considering spin, as it does not couple to the conformal generators.} Thus in Section \ref{sec:MassNotZero}, we consider the tensor product of two generic massive representations
\begin{equation}
    \mathcal{L}_{\Delta_1,m_1} \otimes \mathcal{L}_{\Delta_2,-m_2}
        \quad\text{with}\quad
        m_1 - m_2 \neq 0\,,
\end{equation}
which describes a majority of $\calH\otimes \calH^*$. Then we devote the entirety of Section \ref{sec:Pyramids} to tensor products forming observables, i.e.
\begin{equation}
    \mathcal{L}_{\Delta_1,m} \otimes \mathcal{L}_{\Delta_2,-m}\,.
\end{equation}

\subsection{\texorpdfstring{$\mathcal{L}_{\Delta_1,m_1} \otimes \mathcal{L}_{\Delta_2,-m_2}$}{LD1m1 x LD2m2} -- Generic Composite Operators}\label{sec:MassNotZero}
The simplest case to consider is the tensor product of two scalar primary representations $\calL_{\Delta_1,m_1}$ and $\calL_{\Delta_2,-m_2}$ with $m_1 - m_2 \neq 0$, i.e.
\begin{equation}\label{eq:massiveTensor}
    \calL_{\Delta_1,m_1} \otimes \calL_{\Delta_2,-m_2}\,.
\end{equation}
We also take $\Delta_i > d/2$ so that there are no singular vectors, and thus $\calL_{\Delta_i,m_i} \cong \calV_{\Delta_i,m_i}$. 

We write a state in $\calL_{\Delta_i,m_i}$ as
\begin{equation}
    \ket*{\Delta_i,m_i;n_{i,0},n_{i,1}} := P_0^{n_{i,0}} P_1^{n_{i,1}} \ket{\Delta_i,m_i}\,.
\end{equation}
Then a basis for \eqref{eq:massiveTensor} is given by
\begin{equation}
    \ket*{n_{1,0},n_{1,1},n_{2,0},n_{2,1}} := \ket*{\Delta_1,m_1;n_{1,0},n_{1,1}} \otimes \ket*{\Delta_2,-m_2;n_{2,0},n_{2,1}}\,,
\end{equation}
with $D$ and $M$ eigenvalues
\begin{align}
   \Delta 
        &:= \Delta_1+\Delta_2 + 2n_{1,0} + 2n_{2,0} + n_{1,1} + n_{2,1}\,,\\
    m
        &:= m_1 - m_2 \,.
\end{align}
We call $k := \Delta - \Delta_1 - \Delta_2$ the level. 

The tensor product \eqref{eq:massiveTensor} is very reducible. In general, we have a decomposition
\begin{equation}\label{eq:massiveTensor2}
    \calL_{\Delta_1,m_1} \otimes \calL_{\Delta_2,-m_2} 
        = \bigoplus_{\ell=0}^{\infty} n_{\ell} \calL_{\Delta_1 + \Delta_2 + \ell, m_1-m_2}\,.
\end{equation}
It is a straightforward exercise, e.g. mimicking \cite{Fitzpatrick:2011dm} (see also \cite{Penedones:2010ue, Kulp:2024scx}), to construct primary states in the tensor product. At level 0 and level 1 we have the primaries
\begin{align}
    \text{level 0}:\quad
        &\qquad\qquad\ket{0,0,0,0}\,,\\
    \text{level 1}:\quad
        &m_1\ket{0,0,0,1}-m_2\ket{0,1,0,0}\,.
\end{align}
At level $2$ there are two different primaries we can construct, which we write as a general linear combination ($a,b\in \bbC$):
\begin{align}
    \begin{split}
    \text{level 2}:\quad 
        & (m_1 \Delta_2 + m_2 \Delta_1) \,
        (a \ket*{0,0,0,2} + b \ket*{0,2,0,0})
        \\
    &- \left(a\, m_2 (2 \Delta_2 - 1) + b\, m_1 (2 \Delta_1 - 1)\right)
        \ket*{0,1,0,1}
        \\
    &+ \left(a\, m_2^2 (2 \Delta_2 - 1)
        - b\, m_1 (2 m_1 \Delta_2 + m_2)\right)
        \ket*{1,0,0,0}
        \\
    &+ \left(b\, m_1^2 (2 \Delta_1 - 1)
        - a\, m_2 (2 m_2 \Delta_1 + m_1)\right)
        \ket*{0,0,1,0}\,.
    \end{split}
\end{align}
In higher dimensions, these primaries would be further broken up into $\mathfrak{so}(d)$ multiplets, etc.

Let us find the precise multiplicity $n_\ell$ appearing in \eqref{eq:massiveTensor2}. First, let us denote the dimension of $\calL_{\Delta,m}$ at level $k$ as $d_{\Delta,m,k} =: d_k$ (it is independent of $\Delta$ and $m$ when there are no null states), and we declare each $d_{<0} = 0$. At level $k$, we have the recursion relation
\begin{equation}
    d_{k+3} = d_{k+2} + d_{k+1} - d_{k}\,.
\end{equation}
We can solve this with the initial conditions $(d_{-2}, d_{-1}, d_0)=(0,0,1)$ to get
\begin{equation}\label{eq:statesAtLevel}
    d_k := \dim (\calL_{\Delta,m})_k = \frac{1}{4}(3+2k+(-1)^k) = \floor{k/2} +1 \,.
\end{equation}
Since all of the summands in \eqref{eq:massiveTensor2} are massive and above the unitarity bound, we can easily count the dimension of both sides of \eqref{eq:massiveTensor2} at each level. The dimension of the LHS of \eqref{eq:massiveTensor2} at level $k$ is
\begin{equation}
    \mathrm{dim}(\calL_{\Delta_1,m_1}\otimes\calL_{\Delta_2,-m_2})_k = \sum_{\ell=0}^k d_{k-\ell} d_\ell
\end{equation}
The dimension of the RHS at level $k$ is just $\sum_{\ell=0}^k n_\ell d_{k-\ell}$. Setting equal and solving for $n_\ell$, we find that
\begin{equation}
    n_\ell = \floor{\ell/2}+1\,.
\end{equation}

\subsubsection{A Useful Basis Change}\label{sec:niceBasis}
It is useful, for explicit constructions, to change bases in the previous problem. We introduce the anti-diagonal combinations:
\begin{equation}
    \tilde{Q} := Q \otimes \mathds{1} - \mathds{1} \otimes Q\,.
\end{equation}
E.g., $\tilde{P}_0$ acts by
\begin{equation}
    \tilde{P}_0 \ket{n_{1,0}, n_{1,1}, n_{2,0}, n_{2,1}} = \ket{n_{1,0}+1, n_{1,1}, n_{2,0}, n_{2,1}} - \ket{n_{1,0}, n_{1,1}, n_{2,0}+1, n_{2,1}}\,.
\end{equation}
Our previous states can be re-written
\begin{equation}\label{eq:antiDiagonal}
    \ket{n_{1,0}, n_{1,1}, n_{2,0}, n_{2,1}} = 2^{-n} (P_0 + \tilde{P}_0)^{n_{1,0}}
    (P_1 + \tilde{P}_1)^{n_{1,1}}
    (P_0 - \tilde{P}_0)^{n_{2,0}}
    (P_1 - \tilde{P}_1)^{n_{2,1}}\ket{0,0,0,0}\,,
\end{equation}
where $n := n_{1,0} + n_{1,1} + n_{2,0} + n_{2,1}$.

The diagonal $P_0$, $P_1$ and anti-diagonal $\tilde{P}_0$, $\tilde{P}_1$ combinations form a much more convenient mutually commuting set of derivative operators to uniquely construct a state at any level. In this notation, we use the basis states
\begin{equation}
    \ket{n_0, n_1, \tilde{n}_0, \tilde{n}_1}_\sim
        := P_0^{n_0} P_1^{n_1} \tilde{P}_0^{\tilde{n}_0} \tilde{P}_1^{\tilde{n}_1} \ket{\Delta_1, m_1} \otimes \ket{\Delta_2, -m_2}\,.
\end{equation}
Such combinations also give a neat presentation of states as polynomials in commuting variables. Consequently, we can realize $K$ and $C$ as differential operators on the space of polynomials. At level $\ell$ this can be used to efficiently construct the $\floor{\ell/2} + 1$ primaries of the schematic form:
\begin{equation}
    \ket{0,0,\tilde{n}_0, \tilde{n}_1}_{\sim} 
    \,\sim\, 
    \calO_1^\dagger \!\overset{\leftrightarrow}{\partial}_{0}^{\raisebox{-1.2ex}{$\scriptstyle \tilde{n}_0$}}\!
    \overset{\leftrightarrow}{\partial}_{1}^{\raisebox{-1.2ex}{$\scriptstyle \tilde{n}_1$}} \!\calO_2\,,
\end{equation}
as well as find Clebsch-Gordan coefficients, see Appendix \ref{sec:explicitPyramid} for more details.

\section{Exploring the \texorpdfstring{$M=0$}{M=0} Sector}\label{sec:Pyramids}
Now we turn to the operators and representations of the ``non-genuine $M=0$ sector.'' The non-genuine $M=0$ composite local operators $\calO_1^\dagger\!\calO_2$ transform under the adjoint action with Leibniz rule (because the OPE is regular), i.e.
\begin{equation}
    [Q,\!(\calO_1^\dagger\!\calO_2)(0)] 
    = ([Q,\!\calO_1^\dagger]\!\calO_2)(0) 
    + (\calO_1^\dagger[Q,\!\calO_2])(0) \,.
\end{equation}
These are dual to states in the TFD with zero total mass charge and appearing in the tensor products $\calL_{\Delta_1,m} \otimes \calL_{\Delta_2,-m}$.

These operators/states have a richer structure because spatial translations $P_i$ and boosts $K_j$ commute, and also because there are no constraints from HT unitary. As before, we will specialize to ($1+1$)-dimensions to avoid spacetime index cluttering, but the following arguments hold in general dimensions with tensorial/spin indices restored.

Our claim is that the tensor product decomposes into a direct sum of reducible but indecomposable ``pyramid representations'' $\calS_{\Delta}$:
\begin{equation}\label{eq:tensorDecomp}
    \calL_{\Delta_1,m} \otimes \calL_{\Delta_2,-m} = \bigoplus_{\ell} p_{\ell} \calS_{\Delta_1+\Delta_2+2\ell}\,,
\end{equation}
initially noticed in \cite{Bekaert:2011qd, Golkar:2014mwa, Chen:2021xkw, Chen:2023pqf}. We give some brief intuition for what makes these representations interesting in Section \ref{sec:Intuition}, then give a general abstract definition in Section \ref{sec:rigorous}. In Sections \ref{sec:conservation} and \ref{sec:TFD} we comment on some consequences for symmetry currents and thermal physics respectively. Some additional technical comments are in Appendix \ref{sec:PyramidConstructive}.

\subsection{Alien States and a First Pass at Pyramids}\label{sec:Intuition}
\begin{figure}[t]
\centering
    \begin{minipage}[b]{0.48\textwidth}
    \centering
\begin{tikzcd}[cramped, column sep=-1.0em, row sep = scriptsize]
	&& {\hphantom{{}_{\sim}}\ket{0,0,0,0}_{\sim}} && \\
	& {\hphantom{{}_{\sim}}\ket{0,1,0,0}_{\sim}} && {\hphantom{{}_{\sim}}\ket{0,0,0,1}_{\sim}} \\
	{\hphantom{{}_{\sim}}\ket{0,2,0,0}_{\sim}} & {} & \begin{array}{c} \begin{matrix}\hphantom{{}_{\sim}}\ket{1,0,0,0}_{\sim}\\\hphantom{{}_{\sim}}\ket{0,1,0,1}_{\sim}\end{matrix} \end{array} & {} & {\hphantom{{}_{\sim}}\ket{0,0,0,2}_{\sim}}
	\arrow["{P_i}"', from=1-3, to=2-2]
	\arrow["{P_0}", from=1-3, to=3-3]
	\arrow["{P_i}"', from=2-2, to=3-1]
	\arrow[no head, from=2-2, to=3-2]
	\arrow["{K_i^{\vphantom{\Sigma}}}"', from=2-4, to=1-3]
	\arrow["{P_i}"',from=2-4, to=3-3]
	\arrow[no head, from=2-4, to=3-4]
	\arrow["{\,\,K_i^{\vphantom{\Sigma}}}"', from=3-3, to=2-2]
	\arrow["{K_i^{\vphantom{\Sigma}}}"', from=3-5, to=2-4]
\end{tikzcd}
    \end{minipage}\hfill
    \begin{minipage}{0.24\textwidth}
    \centering
\begin{alignat*}{2}
        &\ket{0,0,0,0}_\sim 
            &&\,\longleftrightarrow\,\, \calO_1^\dagger\!\calO_2\\
        &\ket{0,n,0,0}_\sim 
            &&\,\longleftrightarrow\,\, \partial_i^n(\calO_1^\dagger\!\calO_2)\\
        &\ket{1,0,0,0}_\sim 
            &&\,\longleftrightarrow\,\, \partial_0(\calO_1^\dagger\!\calO_2)\\
        &\ket{0,0,0,\tilde{n}_1}_\sim 
            &&\,\longleftrightarrow\,\,
            \mathcal{O}_1^\dagger \overset{\leftrightarrow}{\partial}_i^{\raisebox{-1.2ex}{$\scriptstyle \tilde{n}_1$}}\mathcal{O}_2\\
         &   &&\quad\,\,\vdots
    \end{alignat*}
    \end{minipage}
    \caption{Left, the states in the tensor product begin to arrange themselves into ``pyramid'' shapes, which are neither highest weight nor lowest weight representations. Primary-descendants form one side of the pyramid, while ``alien states'' form the other side. Right, the operators dual to the first few states in the pyramid.}
    \label{fig:pyramidRep}
\end{figure}
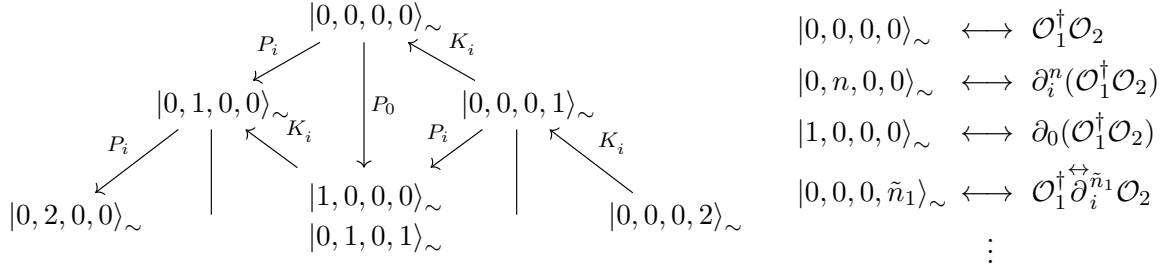

For intuition, we start by considering some states in the tensor product $\calL_{\Delta_1,m} \otimes \calL_{\Delta_2,-m}$ in the notation of Section \ref{sec:niceBasis}. This situation is somewhat more complex than the previous examples for two reasons: first, there are states which are algebraic primary-descendants, and second, there are states which are neither primary nor descendant.
\begin{enumerate}
    \item \textbf{Primary-Descendants.} Consider the tensor product state
    \begin{equation}
        \ket{0,0,0,0}_\sim := \ket{\Delta_1,m}\otimes\ket{\Delta_2,-m}
    \end{equation}
    dual to $\calO_1^\dagger\!\calO_2$. It is clearly primary under the diagonal action of the Schr\"odinger algebra. Given that, there also exists an infinite tower of primary-descendants of the form
    \begin{equation}
        \ket{0,n_1,0,0}_\sim = P_1^{n_1} \!\ket{0,0,0,0}_\sim\,.
    \end{equation}
    For physical states in the HT Hilbert space $\calH$, similar states were modded out by unitarity constraints. Here we are not forced to make such restrictions a priori. More generally, given any $M=0$ Schr\"odinger primary state $\ket{\psi}$, the states $P_1^{n_1} \!\ket{\psi}$ will be primary-descendants of $\ket{\psi}$.
    \item \textbf{Alien States.} The states which are neither primaries nor descendants are more exotic. These include states of the form
    \begin{equation}\label{eq:theAlien}
        \ket{0,0,0,\tilde{n}_1}_\sim = \tilde{P}_1^{\tilde{n}_1}\!\ket{0,0,0,0}_\sim\,.
    \end{equation}
    Clearly this is not the descendant of any state (it has no diagonal derivatives), and it is not a Schr\"odinger primary,\footnote{It is an $\mathfrak{sl}(2,\bbR)$ primary, which could be useful for 1d reductions or special kinematic configurations \cite{Pal:2016rpz, Pal:2018idc}.} instead
    \begin{equation}
        C_0\ket{0,0,0,\tilde{n}_1}_\sim = 0 \,,\quad
        K_1\ket{0,0,0,\tilde{n}_1}_\sim = 2mn \ket{0,0,0,\tilde{n}_1\!-\!1}_\sim\,.
    \end{equation}
    Such states form typical examples of ``alien states'' (dually ``alien operators'') following \cite{Golkar:2014mwa}. In addition to these alien states, there are also states which are descendants of aliens, and so on. Examples of alien operators include the probability current density $j_i(x)$ and spatial stress tensor $T_{ij}(x)$, which become the number density $n(x)$ when lowered. We illustrate this in Figure \ref{fig:pyramidRep}.
\end{enumerate}

Motivated by this, consider the tensor product $\calL_{\Delta_1,m}\otimes \calL_{\Delta_2,-m}$, and let $\ket{\psi}$ be a Schr\"odinger primary vector, we define
\begin{equation}
    \calR_{\psi} :=  \bbC \{P_0^{n_0}P_1^{n_1}\tilde{P}_1^{\tilde{n}_1}\!\ket{\psi} \,|\,  n_0,n_1,\tilde{n}_1 \geq 0 \}\,.
\end{equation}
$\calR_{\psi}$ contains all of the descendants of $\ket{\psi}$ and some alien-type operators ascending to $\ket{\psi}$. In general, $\calR_\psi$ is not a Schr\"odinger module, as it is not $C_0$-closed, unless $\tilde{K}_1\! \ket{\psi} \in \calR_{\psi}$. Instead, it is only a module for the diagonal Lifshitz subalgebra.

One nice simplification occurs if we demand $\ket{\psi}$ is a vector with lowest weight in $\calR_{\psi}$.\footnote{We say vector with lowest weight, because this is not necessarily a lowest-weight module where every vector is obtained by acting on this state with lowering operator. It simply has the lowest $D$ grading.} In this case, $\tilde{K}_1 \!\ket{\psi} = 0$ by assumption and $\calR_{\psi}$ closes into a Schr\"odinger module. Thus, given a lowest weight Schr\"odinger primary vector $\ket{\psi}$ of dimension $\Delta$, mass $0$, and annihilated by $\tilde{K}_1$ in the tensor product, we define the \textit{pyramid module} $\calS_{\psi}$ to be all descendants built from $\ket{\psi}$ and its alien operators, i.e.,\footnote{The $\tilde{K}_1$ condition is not particularly satisfying, in part because it depends on a particular embedding into a tensor product module; we will give an intrinsic definition in the next section.}
\begin{equation}
    \calS_{\psi} := \bbC \{P_0^{n_0}P_1^{n_1}\tilde{P}_1^{\tilde{n}_1}\!\ket{\psi} \,|\,  n_0,n_1,\tilde{n}_1 \geq 0,\, C_0\! \ket{\psi} = K_1\! \ket{\psi} = \tilde{K}_1\! \ket{\psi} = 0\}\,.
\end{equation}

As mentioned, the tensor product \eqref{eq:tensorDecomp} decomposes over a direct sum of pyramid representations. We can see the pattern emerge in $\calL_{\Delta_1,m} \otimes \calL_{\Delta_2,-m}$ with a few simple examples:
\begin{enumerate}\setlength\itemsep{0em}
    \item[0.] \textbf{Level 0}. Here we only have the Schr\"odinger primary
    \begin{equation}
        \ket{\psi_0} := \ket{0,0,0,0}_\sim\,.
    \end{equation}
    The state $\ket{\psi_0}$ is annihilated by $C_0$, $K_1$ and $\tilde{K}_1$, and so forms the top of a pyramid module $\calS_{\psi_0} \cong \calS_{\Delta_1+\Delta_2}$, shown in Figure \ref{fig:pyramidRep}.
    \item[1.] \textbf{Level 1}. The only conformal primary operator is the primary-descendant
    \begin{equation}
        \ket{\psi_1} := \ket{0,1,0,0}_\sim\,,
    \end{equation}
    which satisfies $\tilde{K}_1\!\ket*{\psi_1} = 2m\! \ket*{\psi_1} \neq 0$, and thus does not correspond to a pyramid. Indeed, it can already be seen inside Figure \ref{fig:pyramidRep}.
    \item[2.] \textbf{Level 2}. There are two linearly independent conformal primaries we can consider
    \begin{align}
        \ket{\psi_{2,1}} 
            &:= \ket{0,2,0,0}_\sim\,,\\
        \ket{\psi_{2,2}} 
            &:= 4m\Delta_- \ket{1,0,0,0}_\sim 
            - 4m(\Delta_+ - 1)\ket{0,0,1,0}_\sim\nonumber\\
            &+ (\Delta_+ - 1)\ket {0,2,0,0}_\sim  
            - 2\Delta_- \ket{0, 1, 0, 1}_\sim
            + (\Delta_+ - 1)\ket{0,0,0,2}_\sim\,.\label{eq:firstPyramidTip}
    \end{align}
    The first $\ket{\psi_{2,1}}$ is the primary descendant $P_1^2\!\ket{\psi_0}$, while the second $\ket{\psi_{2,2}}$ is a new primary operator. We note that $\tilde{K}_1\!\ket{\psi_{2,2}} = 0$, and thus it generates its own pyramid $\calS_{\psi_{2,2}}$ different from the one in Figure \ref{fig:pyramidRep}.
\end{enumerate}
We see the general pattern, schematically: anti-symmetric spatial derivatives $\tilde{P}_1$ correspond to alien operators, and anti-symmetric time derivatives $\tilde{P}_0$ correspond to different pyramid representations. Thus we have each $p_\ell = 1$ and
\begin{equation}
    \calL_{\Delta_1,m} \otimes \calL_{\Delta_2,-m} = \bigoplus_{\ell} \calS_{\Delta_1 + \Delta_2 + 2\ell}\,.
\end{equation}
A general formula for the operators satisfying these conditions (tops of pyramids) is given in Appendix \ref{sec:explicitPyramid}. Dimensionally, we compute
\begin{equation}\label{eq:dimS}
    \dim (\calS_{\Delta})_k = \floor{\frac{(k+2)^2}{4}}\,.
\end{equation}

All of this discussion was true for generic $\Delta$. When one (or both) $\Delta_i =  d/2$, the module $\calL_{\Delta_i,m}$ is different from $\calV_{\Delta_i,m}$. In practice, this leads to some identifications within the pyramid. For example, when both $\Delta_i = d/2$, then
\begin{equation}
    \ket{1,0,0,0}_\sim = \frac{1}{2m}\ket{0,1,0,1}_\sim\,,
\end{equation}
thus $\ket{1,0,0,0}_{\sim}$ and $\ket{0,1,0,1}_{\sim}$ are identified in the middle node of Figure \ref{fig:pyramidRep}.

\subsection{Pyramids Built by Aliens}\label{sec:rigorous}
We can now give an intrinsic definition of a pyramid module $\calS_\Delta$ of weight $\Delta$: starting with $\calS_0 := \calV_{\Delta,0}$, we define a series of extensions by Short Exact Sequences (SESs)
\begin{equation}\label{eq:SES}
    0 
        \to \calS_{k-1}
        \to \calS_{k}
        \to \calV_{\Delta+k,0} 
        \to 0\,,\qquad k\geq 1\,.
\end{equation}
The pyramid $\calS_{\Delta}$ is defined to be the direct limit of this sequence with additional constraints.

The SES requires additional information about the gluing of $\calV_{\Delta+k,0}$ to $\calS_{k-1}$, i.e., the extension class. By construction, any SES of the form \eqref{eq:SES} has $\calS_k/\calS_{k-1} \cong \calV_{\Delta+k,0}$, and at each step there is an equivalence class $[v_k] \in \calS_k/\calS_{k-1}$ corresponding to the lowest weight vector in $\calV_{\Delta+k,0}$, satisfying
\begin{equation}
    D[v_k] = (\Delta+k) [v_k]\,,\quad
    M[v_k] = [0]\,,\quad
    C_0[v_k] = [0]\,,\quad
    K_1[v_k] = [0]\,.
\end{equation}
To define a pyramid module, we further demand the existence of a special lift $v_k$ of each class $[v_k]$. In particular, for each $[v_k]$, we demand that there exists a lift $v_k \in \calS_k$ satisfying:\footnote{We could actually demand slightly less generally that $K_1 v_k = \alpha_k v_{k-1}$ for some $\alpha_k$, but we can simply rescale the vectors so that this coefficient is $1$. In the pyramids realized as submodules in tensor products $\calL_{\Delta,m} \otimes \calL_{\Delta,-m}$, a natural normalization condition is often $\alpha_k = 2m$ because $[K_1,\tilde{P}_1] = 2m$.}
\begin{equation}
    D v_k = (\Delta+k)v_k\,,\quad
    Mv_k = 0\,,\quad
    C_0 v_k = 0\,,\quad 
    K_1v_k = v_{k-1}\,.
\end{equation}
In the intuition of Section \ref{sec:Intuition}, this is exactly the addition of alien operators/states along the right-hand side of the pyramid in Figure \ref{fig:pyramidRep}.

However, it is possible that there is more than one lift of $[v_k]$ satisfying the conditions, say $v_k$ and $\tilde{v}_k$. By definition, the difference
\begin{equation}
    w_k := v_k - \tilde{v}_k \in \calS_{k-1}
\end{equation}
satisfies
\begin{equation}
    Dw_k = (\Delta+k) w_k\,,\quad
    Mw_k = 0\,,\quad
    C_0w_k = 0\,,\quad
    K_1w_k = 0\,.
\end{equation}
This is a massless primary vector in $\calS_{k-1}$. We already saw in  Section \ref{sec:Intuition} that, in addition to the primary operator $v_0$ at the top of a pyramid, there are also infinitely many of its primary descendants $P_1^{n_1} v_0$ (in tensor product notation, the top object was extra special for satisfying $\tilde{K}_1 v_0 = 0$, but that has no meaning here). To define a pyramid representation abstractly, all primaries should be the top of the pyramid or its primary descendant.\footnote{Said differently, the restriction that all primaries are the top of a pyramid or its descendants avoids the introduction of ``isolated primaries'' forming the tops of their own pyramid representations.} We can enforce this abstractly by demanding:
\begin{equation}
    \dim \Ker (C_0 \oplus K_1)_{k} = 1\quad \text{for all $k$.}
\end{equation}
This effectively declares that alien operators $v_k$ are defined up to gauge transformations
\begin{equation}
    v_k \sim v_k + \alpha P_1^k v_0\,,
\end{equation}
and that all gauge transformations are of this type.

\vskip 0.5cm
\noindent Some comments are in order:
\begin{itemize}\setlength\itemsep{0em}
    \item The pyramid module $\calS_{\Delta}$ is a reducible but indecomposable Schr\"odinger module, with infinitely many Schr\"odinger submodules, described by the filtration:
    \begin{equation}\label{eq:Filtration}
    \calV_{\Delta,0} 
        = \calS_{0} 
        \subset \calS_{1}
        \subset \dots
        \subset \calS_{\infty} = \calS_\Delta\,.
    \end{equation}
    The filtration degree $N$ could be called the ``alien number.''
    \item In the SES, the number of states in $(\calS_{\Delta})_k$ is
    \begin{equation}\label{eq:dimS2}
        \dim (\calS_{\Delta})_k = \sum_{\ell = 0}^k \dim(V_{\Delta+\ell,0})_{k-\ell} = \sum_{\ell = 0}^k \left(\floor{\frac{k - \ell}{2}}+1\right) = \floor{\frac{(k+2)^2}{4}}\,.
    \end{equation}
    This matches our bottom-up construction in \eqref{eq:dimS}.
    \item The pyramid module is also a maximal module in the following sense: the Schr\"odinger OPE implies that $K_i$ should be surjective (see Appendix \ref{sec:PyramidConstructive}), so every element of $(\calS_\Delta)_k$ should be the $K_i$-image of something in $(\calS_{\Delta})_{k+1}$. The definition of a pyramid as a direct limit of SESs ensures $K_i$ is surjective by extension ``all the way to the right'' in Figure \ref{fig:pyramidRep}. As it turns out, if we try to add more things at level $k+1$, we will end up adding an ``isolated primary'' which starts its own pyramid/module and spoiling indecomposability.
    \item For special $\Delta$, additional redundancies must be removed from the preceding discussion due to shortening conditions. For example, when $\Delta = 1$, there is an additional gauge redundant state at level-2 generated by $w_2' = P_1 v_1 - P_0 v_0$ which satisfies the conditions of $w_N$ but is not $P_1^2 v_0$. This is obviously the same shortening condition that we saw in the tensor product decomposition, and is related to the fact that, for non-generic $\Delta$, we must take suitable quotients.
\end{itemize}

\subsection{Example: Currents, Conservation Laws, and the Space of Theories} \label{sec:conservation}
We can consider the role played by conserved currents in Schr\"odinger CFTs. While constraints from unitarity bounds are not as strong as the relativistic case, the conservation equations become more powerful when embedded in a pyramid module.

In a non-relativistic CFT, a conserved current consists of an $\mathfrak{so}(d)$ vector $J^i$ and a singlet $J^0$ satisfying
\begin{equation}\label{eq:currentCons}
    P_0 J^0 + P_i J^i = 0\,.
\end{equation}
We assume the conserved current integrates to a charge $Q_J$ on a spatial slice and defines a good quantum number, commuting with both $D$ and $M$, then
\begin{equation}
    \Delta_{J^0} = d
    \quad\text{and}\quad
    \Delta_{J^i} = d+1\,,
\end{equation}
and both are massless. 

The massless current constraint is extremely strong and raises some interesting questions about the space of interacting Schr\"odinger CFTs. In particular, if there are no non-trivial $M=0$ states in the harmonic trap spectrum, then $J^0$ is necessarily a composite ``non-genuine'' operator and is right at the non-genuine unitarity bound. i.e., it is a quadratic composite made from a creation operator $\calO_1^\dagger$ and an annihilation operator $\calO_2$, which are themselves at the unitarity bound $\Delta_1 = \Delta_2 = d/2$. We do not expect any local operators to be exactly at the unitarity bound except the fundamental fields of free theories or those defined by perturbation theory from a free theory. This argument also applies to the $M$ current itself. Our expectation is thus that interacting Schr\"odinger CFTs with global symmetry: are approximately free, or have genuine massless operators, or have no currents for symmetry charges.

In any case, let us assume such a current exists. Then by applying $C_0$ to both sides of \eqref{eq:currentCons}, we have
\begin{equation}
    0 = DJ^0 + K_i J^i + P_\mu (C_0 J)^\mu\,.
\end{equation}
Unitarity constrains both $J^0$ and $J^i$ from being primary without the current becoming essentially trivial.\footnote{If both are Schr\"odinger primary, then $D J^0 = 0$. A dimensionless current violates the unitarity bound, so $J^0 \equiv 0$ and $P_i J^i = 0$. Then using successive $K_j$ on the surviving conservation equation renders $J^i$ uninteresting.} Instead, a natural choice is to demand that currents remain $\mathfrak{sl}(2,\bbR)$ primaries, then
\begin{equation}
    0 = \Delta_{J^0} J^0 + K_i J^i\,.
\end{equation}
In fact, we can consider even more general conserved current pairs
\begin{equation}
   P_0 J^{0i_1\cdots i_k} + P_i J^{ii_1\cdots i_k} = 0\,.
\end{equation}
So long as both components are $\mathfrak{sl}(2,\bbR)$ primaries, then applying $C_0$ gives
\begin{equation}
    K_iJ^{ii_1\cdots i_k} = - \Delta_{J^{0i_1\cdots i_k}} J^{0i_1\cdots i_k}\,,
\end{equation}
i.e. $J^{0I}$ is one lower alien number than $J^{iI}$ in a pyramid module.

This is what happens in the pyramid for the number density in the free theory. In this case, the number density and probability currents are
\begin{equation}
    J^0(x) := n(x) = \phi^\dagger\phi(x)
    \quad\text{and}\quad
    J^i := \frac{-1}{2m}(\partial^i\phi^\dagger\phi - \phi^\dagger\partial^i\phi)\,.
\end{equation}
The number density and probability current are the top element $v_0$ and first alien operator $v_1$ in a pyramid representation respectively, i.e. $K_i J^i = -d J^0$. 
In the free theory, this also holds for the stress tensor and probability current, i.e.
\begin{equation}
    J^{0j} := J^j
    \quad\text{and}\quad
    J^{ij} := T^{ij} = \frac{1}{4m^2}\phi^\dagger
            \overset{\leftrightarrow}{\partial}^{\raisebox{-1.2ex}{$\scriptstyle i$}}
            \overset{\leftrightarrow}{\partial}^{\raisebox{-1.2ex}{$\scriptstyle j$}}\!
            \phi\,,
\end{equation}
then 
$K_i J^{ij} = -(d+1) J^{0j}$, and so on.

Better yet, conservation laws in one level of the pyramid can lead to conservation laws at other levels, even in interacting theories, generalizing the results of \cite{Bekaert:2011qd}. To see this, suppose we have current conservation for the ``$N+2$ component tensors'' $(J^{0jI}, J^{ij I})$, then we have
\begin{equation}
    K_j J^{0jI} = - \Delta_{J^{00I}} J^{00I}
    \quad\text{and}\quad
    K_j J^{ijI} = - \Delta_{J^{0iI}} J^{0iI}\,.
\end{equation}
For consistency with scaling dimension, we should have $\Delta_{J^{0iI}} = \Delta_{J^{00I}} +1$. Now we apply $K_j$ to the conservation equation to find
\begin{align}
    0 
         &= K_{j} P_0 J^{0 j I} + P_i K_{j} J^{ij I}\\
       0 &= P_0 J^{00 I} + P_iJ^{0 i I}\,.
\end{align}
Thus the current is conserved for the ``$N+1$ component tensors'' $(J^{00I}, J^{0iI})$.

\subsection{Example: Pyramid Contributions to Thermal Calculations}\label{sec:TFD}
Given $\calH$ with its unitary inner product, we have a unitary product on the double Hilbert space $\calH \otimes \calH^*$ in the obvious way.  $\calH \otimes \calH^*$ carries a left and right Schr\"odinger action, and we use the diagonal and anti-diagonal notation
\begin{equation}
    Q := Q\otimes 1 + 1 \otimes Q
    \quad\text{and}\quad
    \tilde{Q} := Q \otimes 1 - 1 \otimes Q\,.
\end{equation}

Inside this Hilbert space, we consider the state
\begin{equation}
    \ket{\ket{\mathrm{TFD}}}_{\beta,\mu}    := \frac{1}{\sqrt{Z(\beta,\mu)}} \sum_{n} e^{-\frac{\beta}{4}(H - \mu \tilde{M})} \ket{n}\otimes \ket{n^*} \,,
\end{equation}
where 
\begin{equation}
    Z(\beta,\mu) 
        := \Tr_{\calH} e^{-\beta (D_L - \mu M_L)}
\end{equation}
is the partition function of the theory with a chemical potential/mass fugacity. Thermal correlators are obtained by the usual
\begin{equation}
    \expval{\calO}_{\beta,\mu} = {}_{\beta,\mu}\!\bra{\bra{\TFD}}{\!\calO_L\!}\ket{\ket{\TFD}}_{\beta,\mu}\,.
\end{equation}
This can be used to study (semi-)universal properties of Schr\"odinger CFTs, e.g. in hydrodynamics or large-charge EFTs \cite{Bekaert:2011qd, Kravec:2018qnu, Favrod:2018xov, Kravec:2019djc, Hellerman:2023myh, Lee:2026azy, Beane:2026mxi}.

For generic $\beta$ and $\mu$, the TFD state is preserved by the twisted-diagonal Schr\"odinger algebra generators:
\begin{equation}
    e^{-\frac{\beta}{4}(H - \mu \tilde{M})} Q e^{\frac{\beta}{4}(H - \mu \tilde{M})}\,,
\end{equation}
which constrain thermal correlators. In particular, this includes $\tilde{H}$ and $M$. In the zero-temperature $\beta \to \infty$ and chemical potential $\mu \to 0$ limit, the TFD state is just our two-copy HT vacuum
\begin{equation}
    \ket{\ket{\TFD}}_{\infty,0} = \ket{\Omega} \otimes \ket{\Omega^*}\,,
\end{equation}
which appears in our state-operator correspondence, preserved by every generator.

Any TFD state is entirely contained within the $M=0$ subsector of the tensor product Hilbert space $\calH \otimes \calH^*$, and further supported entirely on the tensor products of the form $\calL_{R} \otimes \overline{\calL_{R}} \subset \calH \otimes \calH^*$ for some rep $R$. For example, take a Hilbert space of the form \eqref{eq:Hilb} and work in (1+1)d to avoid spin indices (and take $N_{\Delta,m} \in \{0,1\}$ to avoid trivial but annoying multiplicities), then
\begin{equation}
    \ket{\ket{\TFD}}_{\beta,\mu} = \frac{1}{\sqrt{Z(\beta,\mu)}} \sum_{\Delta,m} \sum_{\ell = 0}^{\infty} e^{-\frac{\beta}{2}(\Delta+\ell-\mu m)} 
    \!\!\!\! \sum_{a \in (\calL_{\Delta,m})_\ell} \!\!\!\!
    \ket{\Delta,m;a} \otimes \ket{\Delta,-m;a^*}
\end{equation}
where $a$ labels the $d_{\Delta,m,\ell}$ normalized states living in $\calL_{\Delta,m}$ at level $\ell$ (recall e.g. \eqref{eq:statesAtLevel}). 

This expansion can be recast in a coupled basis using the pyramid modules. In particular, the TFD state has a decomposition over the pyramid modules as
\begin{align}
    \ket{\ket{\TFD}}_{\beta,\mu} = \frac{1}{\sqrt{Z}} \sum_{\Delta,m}\sum_{\ell = 0}^{\infty} e^{-\frac{\beta}{2}(\Delta-\mu m)} \ket*{\ket*{\widehat{\mathrm{TFD}}}}_{\beta; \Delta,m,\ell}\,,
\end{align}
where $\ket*{\ket*{\widehat{\mathrm{TFD}}}}_{\beta; \Delta,m,\ell} \in \calS_{2\Delta+2\ell}$ is the ``thermal block'' contribution from each pyramid. See Appendix \ref{sec:explicitPyramid} for the detailed form of $\ket*{\ket*{\widehat{\mathrm{TFD}}}}_{\beta; \Delta,m,\ell}$. In this coupled basis we get the exact results for diagonal charges on the thermal state and it would be interesting to use it to probe various thermodynamic properties in future works.

\section{Genuine Massless Modifications and Light-Cone Logarithms}\label{sec:Bootstrap}
When genuine massless operators are in the spectrum, i.e. non-trivial representations $\calL_{\Delta,0}$, many important properties of the non-relativistic CFT change. In particular, there is no longer a non-renormalization theorem ensuring regularity of ``daggered'' and ``undaggered'' OPEs, preventing us from canonically defining composite operators. This is true even in perturbation theory, see \cite{NR0Mass}, where massless particles can now run in virtual loops in the HT vacuum. As we will see, this gives the Schr\"odinger CFT behaviours analogous to usual Lorentzian CFTs.\footnote{In this section, we return to real time conventions. See Appendix \ref{app:algebra}.}

To see this mathematically, we consider a four-point function of scalars and their conjugates (for simplicity), and ordered in time $t_1 > t_2 > t_3 > t_4$ so there is no question of OPE convergence channels
\begin{equation}
    A(x_i) := \mel*{\Omega}{\calO(x_1)\!\calO^\dagger(x_2)\!\calO(x_3)\!\calO^\dagger(x_4)}{\Omega}\,.
\end{equation}
Let us consider the $(12)(34)$ OPE (aka $s$-channel), then we have
\begin{align}
    A(x_i)
        &= \sum_{\ell,r}C_{\calO\!\calO^\dagger\! \ell}(z_{12}, x_{12}, \partial_2) C_{\calO\!\calO^\dagger \!r}(z_{34}, x_{34}, \partial_4)
        \mel*{\Omega}{\calO_\ell(x_2)\!\calO_r(x_4)}{\Omega}\\
        &= \frac{e^{im(z_{12}+z_{34})/2}}{(t_{12} t_{34})^{\Delta_{\calO}}}
        \sum_{g} \frac{(C_{\calO \!\calO^\dagger\! g})^2}{t_{12}^{-\Delta/2}t_{34}^{- \Delta/2}}
        (1+ \calD_g(t_{12})+\dots)(1+ \calD_{g}(t_{34})+\dots) \frac{1}{(t_{24})^{\Delta}}\,.
\end{align}
In the first line, $\ell$ and $r$ sum over genuine massless primaries. In the second line, we have expanded the $C(z,x,\partial)$ into the standard form of a dynamical three-point coefficient, global power of $x$, and sum over descendants $\calD_g$, as well as inserted the massless genuine two-point function. We have taken advantage of the fact that genuine massless operators do not depend on position, which dramatically simplifies the expressions, leaving only dependence on $t$ in many expressions, and turning the three-point coefficient functions $C_{ijk}(z)$ into three-point coefficient constants $C_{ijk}$ \cite{NR0Mass}. We have suppressed $i\epsilon$'s for brevity.

In this case, the OPE and conformal block expansion has effectively reduced to the familiar 1d $SL(2,\bbR)$ block expansion, with a leading exponential prefactor, and a $z = 2$ change of the scaling dimensions.
Thus we have a conformal block expansion of the correlation function
\begin{equation}\label{eq:4ptBlock}
    A(x_i) = \frac{e^{im(z_{12}+z_{34})/2}}{(t_{12}t_{34})^{\Delta_{\calO}}} \sum_g (C_{\calO \!\calO^\dagger\! g})^2 G_{\Delta}(s)\,,
\end{equation}
where
\begin{equation}
    G_{\Delta}(s) := s^{\frac{\Delta}{2}} \,_2F_1\!\left(\tfrac{\Delta}{2},\tfrac{\Delta}{2},\Delta,-s\right) 
\end{equation}
is the (all external same) 4-point conformal block and $s$ is the 1d conformal cross-ratio $s := t_{12}t_{34}/t_{23}t_{14}$.

Now we can consider the expansion of \eqref{eq:4ptBlock} in the crossed $t$-channel regime. This is the limit $t_{23} \to 0$ and/or $s \to \infty$. The limit of an individual $s$-channel conformal block is
\begin{equation}
    G_{\Delta}(s) \xrightarrow[]{s\xrightarrow[]{}\infty} A_{\Delta} \log(s) + B_{\Delta} + O(s^{-1}\log(s))
\end{equation}
for some $\Delta$ dependent constants $A_\Delta$ and $B_\Delta$. In this limit, we can separate the block into a (leading) divergent and regular piece
\begin{equation}
    G_{\Delta}(s) \xrightarrow[]{s\xrightarrow[]{}\infty} - A_\Delta \log(t_{23}) + A_\Delta \log\left(\frac{t_{13}t_{34}}{t_{14}}\right) + B_\Delta + O(s^{-1} \log(s))\,.
\end{equation}
Separating off the 1d conformal cross-ratio $s$ makes these divergences analogous to lightcone divergences (since the operator points may still be far separated in space).

Very naively, these leading terms in the asymptotic expansion would suggest a logarithmic OPE
\begin{equation}
    \calO^\dagger(x_2) \!\calO(x_3) \stackrel{?}{\sim} \log(t_{23}) Q(x_3) + S(x_3)\,,
\end{equation}
for some non-genuine operator $Q$ of conformal dimension $2\Delta_{\calO}$ and its log-partner $S$. However, this logarithm is only one of an infinite collection of log divergences from the blocks. Resumming these logs, we interpret them as a shift of the conformal dimensions of the exchange operator $\Delta \mapsto \Delta + \gamma$ \cite{Simmons-Duffin:2016wlq, NR0Mass}. Thus we have the following result: \textit{genuine massless operators give anomalous dimensions to the non-genuine massless operators in the crossed-channel, and vice-versa}.

Implicit in the last result is the following: if a theory has a finite number of non-trivial genuine massless states, it is logarithmic.  Indeed, any finite number of intermediate OPE channels is not sufficient for renormalization. Conversely, a unitary non-logarithmic CFT requires an infinite number of non-trivial genuine massless states (or none at all). A special case is when there are no genuine massless operators at all, then the 4-point function \eqref{eq:4ptBlock} simplifies to
\begin{equation}
    A(x_i) = \frac{e^{im(z_{12}+z_{34})/2}}{(t_{12}t_{34})^{\Delta_{\calO}}}\,.
\end{equation}
This is called factorization of the four-point function \cite{NR0Mass}. Then in the $t_{23} \to 0$ limit, the cross-channel does not give any contributions at all, and the $\calO(x_2)\calO^\dagger(x_3)$ goes un-renormalized. Consequently, we confirm that factorization implies non-renormalization.

\acknowledgments We would like to thank Philip Argyres, Diego Delmastro, Rajeev Erramilli, Ryan Lanzetta, Ohad Mamroud, Fedor Popov, and Adar Sharon for useful conversations. The work of JK was supported by the NSERC PDF program. The work of MB is supported by the FRQNT doctoral training scholarship.

\appendix

\section{Schr\"odinger Algebra Conventions and the BPZ Inner Product}\label{app:algebra}
In this appendix, we record our conventions for the real-time harmonic trap, and explain the relation to the Euclidean generators in the main text. Since we never use the real-time generators for computations in the main text, no confusion should arise from using the same symbols.

\subsection{Real Time Schr\"odinger Algebra}\label{sec:SchroAlg}
In ($d+1$)-dimensions, the Schr\"odinger algebra is
\begin{equation}\label{eq:LeviMalcev}
    \mathfrak{sch}_d = (\mathfrak{sl}(2,\bbR) \times \mathfrak{so}(d)) \ltimes \mathfrak{h}_{d}\,,
\end{equation}
generated by dilatation $D$, time translation $P_0$, special conformal transformation $C_0$, rotations $M_{ij}$, boosts $K_i$, and spatial translations $P_i$, with a central element $M$. The commutation relations are:
\begin{alignat}{3}
    [D,P_0] 
        &= 2i P_0\,,\quad
    &[C_0,P_0] 
        &= i D\,,\quad
    &[D,C_0] 
        &= -2i C_0\,,\nonumber\\
    [D,P_i] 
        &= i P_i\,,\quad
    &[K_i,P_j] 
        &= i \delta_{ij} M\,,\quad
    &[D,K_i] 
        &= -i K_i\,,\nonumber\\
    [C_0,P_i] 
        &= i K_i\,,\quad
    &&&[P_0,K_i] 
        &= -i P_i\,,\label{eq:SchrodingerAlgebra}\\
    [M_{ij},P_k] 
        &= i(\delta_{ik} P_j-\delta_{jk}P_i)\,,\mkern-36mu
    &&&[M_{ij},K_k]
        &= i(\delta_{ik} K_j-\delta_{jk}K_i)\,,\nonumber
\end{alignat}\vskip -1.8em
\begin{equation}
    \quad[ M_{ij}, M_{kl}]
        = -2i(\delta_{j[l} M_{k]i}-\delta_{i[l} M_{k]j})\nonumber\,.
\end{equation}
Note: $[K_i,P_j]$ is proportional to the central element $M$, and thus commute in the massless sector. Our conventions are chosen so that all the generators are Hermitian, e.g. $P_0^\dagger = P_0$ and $D^\dagger = D$, and generate natural actions in the real-time plane $\bbR^{d+1}$ e.g. $P_0 \sim i\partial_t$.

The natural Hamiltonian in Schr\"odinger CFT is the HT Hamiltonian
\begin{equation}
    H_{\HT} := P_0 +\omega^2 C_0\,,
\end{equation}
where $\omega \in \bbR_{>0}$. $H_{\HT}$ generates time translations in the ``harmonic trap geometry'' $M_{\HT} := (\bbR \times S^0) \times \bbR^{d}$ not the plane $\bbR^{d+1}$. The two geometries are related by a Schr\"odinger-Weyl transformation. Raising and lowering operators for $H_{\HT}$ are \cite{Goldberger:2014hca}
\begin{equation}\label{eq:HRRaiseLower1}
    P_{\pm i}
        = \frac{1}{\sqrt{2\omega}}P_i \pm i \sqrt{\frac{\omega}{2}}K_i\,,\quad
    L_{\pm}
        = \frac{1}{2}\left(\frac{1}{\omega}P_0-\omega C_0 \pm i D\right)\,.
\end{equation}
They satisfy
\begin{alignat}{3}
    [H_{\HT}, P_{\pm i}] 
        &= \pm \omega P_{\pm i}\,,\quad
    &[H_{\HT}, L_{\pm}] 
        &= \pm 2\omega L_{\pm}\,,\quad\\
    [P_{- i}, P_{+ j}] 
        &= \delta_{ij} M\,,\quad
    &[L_{-}, L_{+}] 
        &= \frac{1}{\omega}H_{\HT}\,,\\
    [P_{-i}, L_{+}] 
        &= P_{+i}\,,\quad
    &[L_{-}, P_{+i}] 
        &= P_{-i}\,.
\end{alignat}

\subsection{Euclidean Schr\"odinger Algebra and the BPZ Inner Product}\label{app:EuclideanRadial}
We can Wick rotate and perform a coordinate transform from the real-time Harmonic trap geometry to the Euclidean plane. Under this mapping, the physical real-time $H_{\HT}$ energy is mapped to scaling dimensions of local primary operators in the Euclidean plane, mirroring the relation between the cylinder spectrum and scaling dimensions in relativistic CFT. This coordinate transform is given by a (non-unitary) conjugation, i.e., an Adjoint action of the complexified Schr\"odinger group. When $\omega=1$, this transformation is \cite{NR0Mass}
\begin{equation}
    A \mapsto V A V^{-1} \,,\quad
    V := e^{\frac{\pi}{4}(P_0 - C_0)}\,.
\end{equation}
Under this coordinate transform,
\begin{equation}\label{eq:HTEnergy}
    H_{\HT} \mapsto VH_{\HT}V^{-1} = -i D\,,
\end{equation}
and the HT ladder operators conjugate to:
\begin{equation}\label{eq:ladderMaps}
    P_{+i} \mapsto P_i\,,\quad 
    P_{-i} \mapsto -i K_i\,,\quad
    L_+ \mapsto P_0\,,\quad
    L_- \mapsto -C_0\,.
\end{equation}
Now $-iD$ is related to the physical energy, $P_i$ and $P_0$ are raising operators, and $K_i$ and $C_0$ are lowering operators. 

The HT inner product and adjoint should be pulled back through this map. Given two states, we have an inner product $\braket*{\psi_1}{\psi_2}_E := \braket*{V^{-1}\psi_1}{V^{-1}\psi_2}_{\HT}$. The adjoint for $\braket*{\,\cdot\,}{\,\cdot\,}_E$ is therefore given by $\dagger_E $, where
\begin{equation}
    A^{\dagger_E} := V^2A^{\dagger}V^{-2}\,.
\end{equation}
Of course, the plane generators are not Hermitian under $\dagger_E$, e.g., from \eqref{eq:ladderMaps} we have $P_{0}^{\dagger_E} = - C_0$ and $P_i^{\dagger_E} = i K_i$. 

In the main text, we re-define our generators from those obtained by conjugation above to be more natural for the Euclidean algebra. In particular, we re-define
\begin{equation}
    \widebar{D} = - i D\,,\quad
    \widebar{C}_{0} = -C_{0}\,,\quad
    \widebar{K}_i = -i K_i\,,\quad
    \widebar{M}_{ij} = -i M_{ij}\,.
\end{equation}
All other barred generators are the same as their non-barred counterparts. These new rescaled generators satisfy identical commutation relations to \eqref{eq:SchrodingerAlgebra}, \textit{but with all factors of $i$ dropped}. E.g. $[\widebar{D}, P_0] = 2 P_0$. With these redefinitions, the adjoint $\dagger_E$ becomes the Shapovalov form on $\mathfrak{sch}_d$:
\begin{equation}
    \widebar{D}^{\dagger_E} = \widebar{D}\,,\quad
    \widebar{P}_0^{\dagger_E} = \widebar{C}_0\,,\quad
    \widebar{P}_i^{\dagger_E} = \widebar{K}_i\,,\quad
    \widebar{M}_{ij}^{\dagger_E} = -\widebar{M}_{ij}\,,\quad 
    \widebar{M}^{\dagger_E} = \widebar{M}\,.
\end{equation}
This exactly matches the action of $\dag_{\mathrm{rad.}}$ in radial quantization  in Euclidean relativistic CFT. Since we only ever use the rescaled Euclidean generators and ${\dagger_E}$, \textbf{we drop the bar on generators, write $\braket*{\,\cdot\,}{\,\cdot\,}$ for the inner product, and write $\dagger$ for ${\dagger_E}$ in the main text.} We keep them for the remainder of this appendix.

A local primary operator $\calO^\dagger_a(x)$ or $\calO_b(x)$ is annihilated by $[\widebar{C}_0,\,\cdot\,]$ and $[\widebar{K}_i,\,\cdot\,]$, and has well-defined scaling, mass, and spin-representation, i.e.
\begin{alignat}{2}
    [\widebar{D}, \calO_a^\dagger(0)] &= \Delta_a \!\calO_a^\dagger(0)\,,\quad
    [\widebar{M}, \calO_a^\dagger(0)] &&= m_a \!\calO_a^\dagger(0)\,,\\
    [\widebar{D}, \calO_b(0)] &= \Delta_b \!\calO_b(0)\,,\quad
    [\widebar{M}, \calO_b(0)] &&= -m_b \!\calO_b(0)\,.
\end{alignat}
By \eqref{eq:HTEnergy}, the scaling dimension of states created by acting on the vacuum with local primaries, and their descendants, are the HT energy spectrum \cite{Nishida:2007pj, NR0Mass}. It is important not to conflate the conjugation $\dagger$ on our (generically) non-real creation and annihilation operators with $\dagger_E$, which is the adjoint for our module. As in standard CFT literature, this is distinguished by the position of $\dagger$ before or after the argument of $\calO$, i.e  generically $\calO^\dagger(x) \neq \calO(x)^{\dagger_E}$.

To illustrate this last point, a state in our lowest weight multiplet is created by acting on the HT vacuum $\ket{\Omega}$ with a local operator at the origin
\begin{equation}
    \ket{a} := \calO^\dagger_a(0) \!\ket{\Omega}\,.
\end{equation}
This state is obviously a lowest-weight vector for the Schr\"odinger algebra, with
\begin{equation}
    \widebar{D} \!\ket{a} = \Delta_a \!\ket{a}\,,\quad
    \widebar{M} \!\ket{a} = m_a\! \ket{a}\,,
\end{equation}
and well defined spin. The analogue of BPZ conjugation in this Euclidean radial quantization scheme is
\begin{equation}
    \calO^\dagger_a(t_E,\vec{x})^{\dagger_E} = t_E^{-\Delta_a} \calO_a(1/t_E,\vec{x}/t_E)\,,
\end{equation}
and thus
\begin{equation}
    \bra{a} = (\calO_a^\dagger(0)\!\ket{\Omega})^{\dagger_E} = \lim_{x\to 0} \bra{\Omega}\! \calO^\dagger_a(x)^{\dagger_E} = \lim_{t_E\to\infty} t_{E}^{\Delta_a} \bra{\Omega}\!\calO_a(t_E)\,.
\end{equation}
For example
\begin{equation}
    \braket{b}{a} = \lim_{t_E \to \infty} t_E^{\Delta_b} \mel*{\Omega}{\calO_b(t_E)\calO_a^\dagger(0)}{\Omega}\,,
\end{equation}
exactly as in usual CFT, with adjusted pre-factors for $z=2$ scaling and spin decoupling. Finally, we note that $\bra{a}$ and/or $\calO_b(\infty)$ of course transforms very differently from $\calO_b(0)$.

\section{Pyramid Modules in the Tensor Product}\label{sec:explicitPyramid}
Here we provide more details on how to construct the pyramid modules in the tensor product, and then how to convert between the tensor and coupled basis. We use this to obtain the pyramid thermal contribution $\ket*{\ket*{\widehat{\TFD}}}_{\beta;\Delta,m,\ell}$. 

To construct the pyramid modules explicitly in the tensor product $\calL_{\Delta_1,m_1} \otimes \calL_{\Delta_2,-m_2}$ we recall the fact that the states
\begin{equation}
    P_0^{n_0} P_1^{n_1} \tilde{P}_0^{\tilde{n}_0} \tilde{P}_1^{\tilde{n}_1} 
    \!\ket{\Delta_1,m_1}\otimes \ket{\Delta_2,-m_2}\,,
\end{equation}
can be considered as a polynomial in the commuting variables $u_0$, $u_1$, $\tilde{u}_0$, $\tilde{u}_1$. We will use this notation for the appendix to remind us that everything is commutative algebra. The actions of $K_1$, $\tilde{K}_1$, and $C_0$ can be represented by the differential operators:
\begin{align}
\label{Kdifferential}
\mathcal{K}
    &= \tilde{u}_1\partial_{\tilde{u}_0}
       + u_1\partial_{u_0}
       + m_-\partial_{u_1}
       + m_+\partial_{\tilde{u}_1}\,,\\
\label{Ktildedifferential}
\tilde{\mathcal{K}}
    &= u_1\partial_{\tilde{u}_0}
       + \tilde{u}_1\partial_{u_0}
       + m_+\partial_{u_1}
       + m_-\partial_{\tilde{u}_1}\,,\\
\label{Cdifferential}
\mathcal{C}
    &= \Delta_+\partial_{u_0}
        +\Delta_-\partial_{\tilde{u}_0}
        +\frac{m_-}{2}\bigl(\partial_{\tilde{u}_1}^{2} + \partial_{u_1}^{2}\bigr)
        +m_+\partial_{\tilde{u}_1}\partial_{u_1}\nonumber\\
    &\quad +u_0\partial_{\tilde{u}_0}^{2}
       +u_1\partial_{\tilde{u}_0}\partial_{\tilde{u}_1}
       +\tilde{u}_1\partial_{\tilde{u}_0}\partial_{u_1}
        +2\tilde{u}_0\partial_{\tilde{u}_0}\partial_{u_0}+\tilde{u}_1\partial_{\tilde{u}_1}\partial_{u_0}
       +u_1\partial_{u_1}\partial_{u_0}
       +u_0\partial_{u_0}^{2}\,,
\end{align}
where $\Delta_{\pm} := \Delta_1 \pm \Delta_2$ and  $m_{\pm} := m_1 \pm m_2$. These differential operators can be used to construct explicit descriptions of the primaries, as described in Section \ref{sec:niceBasis}. 

Here, we are interested in primaries that are tops of pyramid representations. In Section \ref{sec:Intuition} we argued that they should correspond to primary vectors which are also annihilated by $\tilde{K}_i$ in the tensor product. We will construct those states now, but also specialize to $\Delta_1 = \Delta_2 = \Delta$ as this is what appears in the TFD state and set $m_+ = 2m$ and $m_- = 0$.

The first step is to find the invariants under $\mathcal{K}$ and $\tilde{\mathcal{K}}$, they are:
\begin{equation}
    A := u_1^2 + \tilde{u}_1^2 - 4m \tilde{u}_0
    \quad\text{and}\quad
    B := u_1 \tilde{u}_1 - 2 mu_0\,.
\end{equation}
They both carry weight $2$ under $D$. The tops of pyramids are therefore of the form
\begin{equation}
    \ket{\psi_k} = f_k(A,B) \ket{\psi_0}\,,
\end{equation}
for some function $f_k(A,B)$. $\ket{\psi_0}$ is the tensor product of the two original lowest weight states, i.e. the top of the first pyramid. Indeed, we note that $A \ket{\psi_0}$ matches with $\ket{\psi_{2,2}}$ from \eqref{eq:firstPyramidTip} after setting $\Delta_- = 0$. 

In order for $\ket{\psi_k}$ to be the top of a pyramid, we must further demand that $f_k(A,B)$ satisfy the $\calC$ differential equation. The corresponding PDE for $f_k(A,B)$ is solvable by integration by parts. However, we only care about polynomial solutions; specifically, $f_k(A,B)$ should be a homogeneous polynomial of total degree $k$ in $A$ and $B$:
\begin{equation}\label{eq:PolynomialAssumption}
    f_k(A,B) = \sum_{\ell = 0}^k c_{k,\ell} A^\ell B^{k-\ell}\,.
\end{equation}
Now a straightforward application of $\calC$ on the polynomial ansatz gives
\begin{equation}\label{eq:answer}
    f_k(A,B) = c_{k,0} \,A^k\, {}_{2}F_{1}\!\left(-\tfrac{k}{2},\tfrac{1-k}{2};2-\Delta-k;\tfrac{4B^2}{A^2}\right)\,.
\end{equation}
All states in the $k$-th pyramid module (of weight $2\Delta + 2k$) are thus of the form:
\begin{equation}
    u_0^{n_0} u_1^{n_1} \tilde{u}_1^{\tilde{n}_1} f_k(A,B) \ket{\psi_0}\,.
\end{equation}

Now we can relate the tensor product basis to the pyramids explicitly. This can be generalized to obtain general Clebsch-Gordan coefficients for the Schr\"odinger group. As in \ref{sec:niceBasis}, we write
\begin{equation}\label{eq:tensorBasis}
    \ket{n_0,n_1,n_0,n_1} = 4^{-(n_0+n_1)}(u_0^2-\tilde{u}_0^2)^{n_0}(u_1^2-\tilde{u}_1^2)^{n_1} \ket{0,0,0,0}\,.
\end{equation}
Intuitively, we know that $\tilde{P}_0$ and/or $\tilde{u}_0$ counts the pyramid module weight in the tensor product, and $\tilde{u}_0$ only appears in $A$, thus to change from tensor products to pyramids, we will simply replace
\begin{equation}
    \tilde{u}_0 = \frac{1}{4m}(u_1^2 + \tilde{u}_1^2-A)\,.
\end{equation}
Setting the normalization $c_{k,0} = 1$, so $A^k$ has coefficient $1$ in \eqref{eq:PolynomialAssumption}, we can also invert\footnote{The appearance of the floor, separating even and odd powers, can be seen by studying the first few terms of \eqref{eq:answer}.}
\begin{equation}\label{eq:AkExpansion}
    A^k = \sum_{\ell=0}^{\floor{ k/2 }}\frac{k!}{(k-2\ell)!\ell!(\Delta+k-2\ell)_\ell} (u_1 \tilde{u}_1 - 2 mu_0)^{2\ell}f_{k-2\ell}(A,B)\,.
\end{equation}
Crucially, this lets us see how $\tilde{u}_0$ powers in the basis \eqref{eq:tensorBasis} become pyramid tops and their descendants, by simply expanding and substituting powers of $A^k$. For example, we expand
\begin{align}
    \ket{0,1,0,1} 
        &= \frac{1}{4}(u_1^2 - \tilde{u}_1^2) \ket{\psi_0}\,,\\
    \ket{1,0,1,0}
        &= -\frac{2(u_1 \tilde{u}_1 - 2 mu_0)^2+\Delta(u_1^2+\tilde{u}_1^2-4mu_0)(u_1^2+\tilde{u}_1^2+4mu_0)}{64m^2\Delta} \ket{\psi_0}\nonumber\\
        &\quad+\frac{u_1^2+\tilde{u}_1^2}{32m^2}\ket{\psi_1} - \frac{1}{64m^2} \ket{\psi_2}\,,
\end{align}
and so on.

Let us call the Clebsch-Gordan coefficient in front of $\ket{\psi_\ell}$ in the tensor product $\ket{n_0,n_1,n_0,n_1}$ as $C_{\Delta,m,\ell}^{n_0,n_1}$, then
\begin{equation}
    \ket*{\ket*{\widehat{\TFD}}}_{\beta,\mu;\Delta,m,\ell} = \sum_{n_0,n_1} e^{-\frac{\beta}{2}(2n_0+n_1)}C^{n_0,n_1}_{\Delta,m,\ell}(u_0,u_1,\tilde{u}_1) \ket{\psi_\ell}_{\Delta,m}
\end{equation}
where $\ket{\psi_\ell}_{\Delta,m}$ is the $\ell$-th pyramid-top in the tensor product of $\ket{\Delta,m}\otimes \ket{\Delta,-m}$. We caution the reader that none of these states are necessarily normalized, so calculations will still require computing the norms $\abs{\ket{n_0,n_1,n_0,n_1}}^2$ or $\abs{\ket{\psi_\ell}}^2$ to proceed.

\section{More Properties of Pyramids}\label{sec:PyramidConstructive}
Here we expand on the pyramid modules $S_{\Delta}$ defined in Section \ref{sec:Pyramids}. We will argue that pyramids are distinguished by the following physical properties:
\begin{enumerate}
    \item[(i)] Bounded indecomposability. The pyramid modules are reducible but indecomposable modules, graded by $D$, with a lowest level subspace. As vector spaces, we have
    \begin{equation}
    \calS_{\Delta} = \bigoplus_{k\geq 0} \calS_{\Delta,k}\qquad
        D(\calS_{\Delta,k}) = (\Delta+k) \calS_{\Delta,k}\,.
    \end{equation}
    This ensures that $D$ remains a diagonalizable operator, that we can continue to raise and lower scaling dimensions, and that scaling dimensions are bounded below. Boundedness from below is reasonable from unitarity bounds of states in the single-copy Hilbert space.
    \item[(ii)] Maximal genericity. All operators compatible with the Schr\"odinger algebra Ward identities appear in the module (it is maximal), and we excise all exceptional $(\Delta,m,\rho)$-dependent shortening conditions between generators in the action of $U(\mathfrak{sch}_d)$.\footnote{For example, when $\Delta_{\mathrm{top}} = d$ there may be some linear redundancies between operators coming from the $d$-dimensional Schr\"odinger equation. These should be removed.}
    \item[(iii)] Surjectivity of $K_i$. We demand that $K_i:\calS_{\Delta,k+1} \to \calS_{\Delta,k}$ is a surjection. As we will explain in Appendix \ref{app:surjOPE}, this demand actually follows from the OPE.
\end{enumerate}
By building in these properties, we can give a bottom-up construction of the pyramid modules. As always, we will work in ($1+1$)d for simplicity, but extensions to higher dimensions follow. We will suppress the $\Delta$ on $\calS_{\Delta}$ going forward. These properties are also satisfied by the SES definition: (i) and (iii) are immediate, while (ii) is less obvious.

\subsection{Arguments for Pyramid Uniqueness}
Our claim is the following: \textit{if a Schr\"odinger module $\calS$ satisfies properties (i), (ii), and (iii), it is a pyramid module.} The proof constitutes the rest of this section.

We start by considering a vector $v_0 \in \calS_0$, it has well-defined scaling and mass and must be primary since no lower levels $\calS_{-k}$ exist in the decomposition of $\calS$, thus
\begin{equation}
    D v_0 = \Delta v_0\,,\quad
    M v_0 = 0\,,\quad
    C_0 v_0 = 0 = K_1 v_0\,.
\end{equation}
Since any vector in $\calS_0$ would have this property, all vectors in $\calS_0$ are primary. Let us call this particular $v_0$ ``the top of the pyramid,'' later we will see that this space was one-dimensional.\footnote{In higher dimensions, the top of the pyramid will actually be some $SO(d)$ spinning-multiplet. This could also be one component of a flavour multiplet, etc.}

Even more generally, the whole module $\calS$ can contain other primary operators. At each level, there is one spatial primary-descendant of $v_0$,
\begin{equation}
    w_k := P_1^k v_0\,.
\end{equation}
All other primaries will be called ``isolated primaries.'' We introduce the (family of) map(s) $\calK_k: \calS_k \to \calS_{k-1} \oplus \calS_{k-2}$ by 
\begin{equation}
    \calK_k(v_k) := (K_1 v_k, C_0 v_k)\,,
\end{equation}
i.e., $\calK = K_1 \oplus C_0$. The primary condition is equivalent to the statement that $v_k \in \Ker \calK_k$, and
\begin{equation}
    \dim \Ker \calK_k = 1 + N_k\,,
\end{equation}
where $N_k$ is the number of isolated primaries at level $k$. 

Now we interrogate $\Im \calK_k$ using the algebraic relations within the module and (ii). First we define a map $\calL_k: \calS_{k-1} \oplus \calS_{k-2} \to \calS_{k-3}$ by
\begin{equation}
    \calL_k(v_{k-1}, v_{k-2}) := C_0 v_{k-1} - K_1 v_{k-2}\,.
\end{equation}
Obviously we have an inclusion $\Im \calK_k \subset \Ker \calL_k$. Property (ii) then implies that $\Im \calK_k = \Ker \calL_k$, i.e., for every pair of operators satisfying $C_0 v_{k-1} = K_1 v_{k-2}$ we have an operator $v_k \in \calS_k$ such that
\begin{equation}
    \calK_k v_k = (K_1 v_k, C_0 v_k) \stackrel{!}{=} (v_{k-1}, v_{k-2})\,.
\end{equation}
This essentially demands that each level of the pyramid is as large as possible. This is rightfully part of the ``genericity'' requirement: if we miraculously find that operators at different levels are equivalent under lowering operators, then there is some operator $v_k$ that explains the origin of that equivalence.

Now it is smooth sailing. By (iii), $K_1$ is surjective, so any operator $v_{k-3} = K_1 v_{k-2}$ for some $v_{k-2}$, and thus is $\calL_{k}(0,-v_{k-2}) = v_{k-3}$. Therefore $\calL_{k}$ surjects onto $\calS_{k-3}$
\begin{equation}
    \Im \calL_{k} = \calS_{k-3}\,.
\end{equation}
Now we can apply rank-nullity to $\calL_k$
\begin{align}
    \dim(\calS_{k-1} \oplus \calS_{k-2})
        &= \dim \Ker \calL_k + \dim \Im  \calL_k\\
        &= \dim \Im \calK_k + \dim \calS_{k-3}\,,
\end{align}
as well as to $\calK_k$
\begin{equation}
    \dim \calS_k = \dim \Ker \calK_k + \dim \Im \calK_k\,,
\end{equation}
to get
\begin{equation}\label{eq:dimSkApp}
    \dim\calS_k = 1 + N_k + \dim \calS_{k-1} + \dim \calS_{k-2} - \dim \calS_{k-3}\,.
\end{equation}

\eqref{eq:dimSkApp} gives us a recursion relation for the dimension, which we can solve:
\begin{equation}\label{eq:dimS3}
    \dim \calS_k 
        = \floor{\frac{(k+2)^2}{4}} + \sum_{\ell=0}^{k} N_\ell \left(\floor{\frac{k-\ell}{2}} + 1\right)\,.
\end{equation}
If all $N_\ell = 0$, the modules $\calS$ built from the (maximal) ascendants of $v_0$, and all of their descendants, by (ii) and (iii), is dimensionally consistent with the bottom-up \eqref{eq:dimS} and top-down SES \eqref{eq:dimS2} module constructions. This is the pyramid built from $v_0$. 

When $N_\ell \neq 0$, we recognize the last term of \eqref{eq:dimS3} from our SES computation
\begin{equation}
    \dim (V_{\Delta + \ell,0})_{k-\ell} = \floor{\frac{k-\ell}{2}} + 1\,.
\end{equation}
Indeed, if we assume that $M_k$ are the number of isolated pyramid tops at level $k$, then $N_k = \sum_{\ell=0}^k M_\ell$ and we can re-write
\begin{equation}
    \dim \calS_k 
        = \floor{\frac{(k+2)^2}{4}} + \sum_{\ell=0}^{k} M_\ell \floor{\frac{(k-\ell+2)^2}{4}} \,.
\end{equation}
From this, it seems dimensionally plausible that the effect of adding isolated primaries is to simply add new pyramids. Thus we anticipate $N_k = 0$ is mandatory from indecomposability (i).

To justify this further, we note that our prior arguments go through without any reference to $v_0$ at all, except to simply name some subset $w_k$ of primary-descendants at level $k$ when studying $\Ker \calK_k$, and implicitly when extending the pyramid so that $K_1$ surjects onto objects in lower levels. If we had chosen one of the primary descendants $w_k$ as our $v_0$, we would have generated a pyramid sub-module of $\calS$, and so on. So any primary builds a pyramid module, and it simply remains to show that any other (non-descendant) primary builds a pyramid which can be disentangled from $\calS$, and so should be ignored by (i).

\subsection{No Isolated Primaries}
Suppose we have a module $\calS$ satisfying properties (ii) and (iii) from before. Let $\{v_0\}$ and $\{q_j\}$ be the set of all (non-descendant) primaries, and let $\calS'$ and $\calQ_j$ be their respective pyramid submodules, which we assume span $\calS$. We will show that the module $\calS$ is reducible, splitting like
\begin{equation}
\label{eq:decomposable}
    \mathcal{S}=\mathcal{S}'\oplus \bigoplus_{j} \mathcal{Q}_j
\end{equation}
as representations. In order to show this, we will show that the pyramid modules $\calQ_j$ generated from the $q_j$ can be ``disentangled'' and split off from the original $\mathcal{S}'$-pyramid built from $v_0$, at most through a change of basis.

Let $\calQ_j$ be an isolated primary module, generated by the ascendants and descendants of the top $q_j$, and define
\begin{equation}
    \calV_j := \calS' + \sum_{\ell\neq j} \calQ_\ell\,.
\end{equation}
We would like to show that $\calQ_j \cap \calV_j = 0$. To this end, let us write
\begin{equation}
    D v_0 = \Delta v_0
    \quad\text{and}\quad
    D q_j = (\Delta+\delta_j) q_j\,,
\end{equation}
and define
\begin{align}
    \calQ_{j,\leq k} := \{w \in \calQ_j \,|\, D w = \lambda w\,, \lambda \leq \Delta + \delta_j + k\}
\end{align}
We call $\calQ_j$ \emph{separable at level $k$} if: 1. $\calQ_{j,\leq k}$ is closed under lowering; and 2. $\calQ_{j, \leq k} \cap \calV_{j} = 0$. This ensures all ascendants and descendants of $q_j$ are not coincidentally in other pyramids, up to level $k$. Then we work inductively, level-by-level, showing that we can redefine our alien operators by ``gauge transformations'' to make $\calQ_j$ separable up to and including level $k$.

So fix some module $\calQ_j$, and let us work inductively in $k$. The $k=0$ base case is trivial, so we will show separability at level $k=1$ (higher levels are the same, but suppressed for brevity). Level $k=1$ of the $\calQ_j$ module is spanned by an ascendant/alien $A_{j,1}$ pre-image of $q_j$ and the primary descendant $P_1q_j$.
\begin{enumerate}\setlength\itemsep{0em}
    \item The new elements added at level one are $A_{j,1}$ and $P_1 q_j$. Under lowering, $\mathfrak{s}_- (P_1 q_j) = 0$ and $K_1 A_{j,1} = q_j \in \calQ_{j,\leq 1}$. The only problematic element is therefore $C_0 A_{j,1} \in \calS_{\delta_j-1} \cap \Ker(K_1)$. Now, because $\calS$ satisfies properties (ii) and (iii), we know that $\calK$ surjects onto $\Ker \calL$, and thus we can find a $v_{\delta_j+1} \in \calS_{\delta_j+1}$ satisfying $\calK(v_{\delta_j+1}) = (0, C_0 A_{j,1})$. Therefore, if we redefine
    \begin{equation}
        A_{j,1} \mapsto A_{j,1}' = A_{j,1} - v_{\delta_j+1}
    \end{equation}
    then $A_{j,1}'$ is the ``correct'' alien lift that ascends to $q_j$ and is annihilated by $C_0$. 
    \item Now we show linear independence. Suppose we have a linear combination $w_{j,1} := \alpha A_{j,1}' + \beta P_1 q_j$, and suppose that it is also in $\calV_j$. From the previous exercise, we know that
    $K_1 w_{j,1} = \alpha q_j$ and $C_0 w_{j,1} = 0$. But $\calV_j$ is closed under $\mathfrak{s}_-$ by its definition as a sum of modules, so this means $\alpha q_j \in \calQ_{j,\leq 0} \cap \calV_{j}$. This intersection is zero by separability at level $k=0$, so $\alpha = 0$ and $w_{j,1} = \beta P_1 q_j$. But $P_1 q_j$ is a primary-descendant of $q_j$ and so is not in the other pyramids.
\end{enumerate}
This completes the induction step from $k=0$ to $k=1$, showing that the $\mathcal{Q}_j$-pyramid is separable at level 1. At higher levels, one must show: if $\mathcal{Q}_j$ is separable at level $k$, the ascendant $A_{j,k+1}$ at level $k+1$ can be chosen/gauge transformed so that $C_0 A_{j,k+1}=0$; and general new linear combinations $\alpha A_{j,k+1}+\beta P_0 w_{j,k-1}+\gamma P_1 w_{j,k}$ are not in $\calV_j$.

\subsection{Surjectivity of \texorpdfstring{$K_i$}{Ki} from OPE}\label{app:surjOPE}
The main difference between the pyramid modules $\calS_{\Delta}$ and standard LWMs $\calL_{\Delta,m}$ are additional ``alien'' operators added at each level (plus descendants for module closure). These additional alien operators can be interpreted as the objects saturating surjectivity of $K_i$, which follows from the OPE, as we now show.

First, we recall that the Schr\"odinger boost generator $K_i$ acts on primaries (in Euclidean conventions) as
\begin{equation}
\label{eq:GalileanBoost}
    [K_i,\mathcal{O}_{\Delta,m}(x)] = i(t\partial_{i}-mx_i)\mathcal{O}_{\Delta,m}(x)\,.
\end{equation}
We also recall that the OPE of two opposite-mass operators takes the form \cite{Goldberger:2014hca},
\begin{equation}\label{eq:OPE}
    \mathcal{O}_{\Delta,m}(x)\mathcal{O}_{\Delta',-m}(0)\sim \sum_{\mathcal{O}\in \mathcal{S}}e^{mz/2} c_{\mathcal{O}}(z) t^{(\Delta_{\mathcal{O}}-\Delta-\Delta')/2} (\calD(x) \cdot \mathcal{O})_{}(0)\,,
\end{equation}
where $z := x^2/t$. For simplicity, we consider scalar primaries and the notation $\calD(x) \cdot \mathcal{O}$ is short-hand for correctly contracting all $SO(d)$/derivative indices of operators in $\calS$ with $x$'s.

Now we argue that consistency of the OPE and non-renormalization imply surjectivity of $K_1$ at each level. The proof proceeds by induction: suppose we know that $K_1$ is surjective up to level $k$ of $\mathcal{S}$, we will show that $K_1$ must be surjective at level $k+1$. The base case holds trivially.

First, we choose a basis $\{\mathcal{O}^k_i\}_{i=1,\dots,|\mathcal{S}_k|}$ of $\mathcal{S}_k$. We let $\mathcal{O}_1^k$ be an alien operator included at level $k$ to ensure surjectivity of $K_1$ by the induction step. This is the only operator at risk of not being in the image $K_1: \mathcal{S}_{k+1} \to \calS_{k}$ at the next level, since all the other operators in $\calS_k$ will have $K_1$-preimages coming from $P_\mu$ descendants. This can be seen by comparing dimension formulas for descendants or by noting that $K_1 P_1$ at each level is of Jordan block form for the generalized eigenvalue $0$. We also choose a basis $\{\mathcal{O}^{k+1}_j\}_{j=1,\dots,|\mathcal{S}_{k+1}|}$ of operators for $\calS_{k+1}$. 

The action of $K_1$ on $\calS_{k+1}$ operators placed at $(t,x)=(0,0)$ can be written as
\begin{equation}
    [K_1,\mathcal{O}_j^{k+1}(0)]=\sum_{i=1}^{|\mathcal{S}_k|} \kappa_{j i} \mathcal{O}^k_i(0)\,,
\end{equation}
for some $\kappa_{ji}$. Applying $K_1$ to both sides of the OPE \eqref{eq:OPE}, and using \eqref{eq:GalileanBoost}, we obtain a recursion relation between three-point coefficients at different levels \cite{Goldberger:2014hca}:
\begin{equation}
    2i\partial_z c_{\mathcal{O}^k_i}(z)=\sum_{j= 1}^{|\mathcal{S}_{k+1}|}  \kappa_{j i}c_{\mathcal{O}^{k+1}_{j}}(z)\,.
\end{equation}

Now suppose, for contradiction, that $K_1$ was not surjective, and that it failed at level $k+1$. As previously explained, the issue must be that $\calO_1^k$ has no $K_1$-preimage. The three-point coefficient recursion relations then imply that:
\begin{equation}
    \partial_z c_{\mathcal{O}_1^N}(z)=0\,.
\end{equation}
However, we recall that in the $\calO_{\Delta,m} \times \calO_{\Delta',-m}$ OPE, there canonically exists a non-genuine non-renormalized operator $\calO_1^{0} := (\calO_{\Delta,m}\!\calO_{\Delta',-m})$ of scaling dimension $\Delta + \Delta'$ as the first leading term in the OPE \cite{NR0Mass}. This operator must have
\begin{equation}
    c_{\calO_1^0}(z) = e^{-m z/2}\,,
\end{equation}
and thus must satisfy $c_{\calO_1^k} = (-im)^k e^{-m z/2} \neq 0$\,. A contradiction. Therefore, compatibility of the OPE, plus the non-renormalization theorems, require $K_1$ to be surjective at each level.

\bibliographystyle{JHEP}
\bibliography{refs}

\providecommand{\href}[2]{#2}\begingroup\raggedright\begin{thebibliography}{10}

\bibitem{NR0Mass}
M.~Boisvert, S.~H. Fadda, J.~Kulp, and R.~M. Yazdi, {\it {Revisiting
  Schr{\"o}dinger CFTs: Factorization, Massless Particles, and a Path to the
  Bootstrap}},  \href{http://arxiv.org/abs/2510.26872}{{\tt arXiv:2510.26872}}.

\bibitem{Henkel:1993sg}
M.~Henkel, {\it {Schrodinger invariance in strongly anisotropic critical
  systems}},  {\em J. Statist. Phys.} {\bf 75} (1994) 1023--1061,
  [\href{http://arxiv.org/abs/hep-th/9310081}{{\tt hep-th/9310081}}].

\bibitem{Henkel:2003pu}
M.~Henkel and J.~Unterberger, {\it {Schrodinger invariance and space-time
  symmetries}},  {\em Nucl. Phys. B} {\bf 660} (2003) 407--435,
  [\href{http://arxiv.org/abs/hep-th/0302187}{{\tt hep-th/0302187}}].

\bibitem{Duval:2024eod}
C.~Duval, M.~Henkel, P.~Horvathy, S.~Rouhani, and P.~Zhang, {\it {Schr\"odinger
  Symmetry: A Historical Review}},  {\em Int. J. Theor. Phys.} {\bf 63} (2024),
  no.~8 184, [\href{http://arxiv.org/abs/2403.20316}{{\tt arXiv:2403.20316}}].

\bibitem{Baiguera:2023fus}
S.~Baiguera, {\it {Aspects of non-relativistic quantum field theories}},  {\em
  Eur. Phys. J. C} {\bf 84} (2024), no.~3 268,
  [\href{http://arxiv.org/abs/2311.00027}{{\tt arXiv:2311.00027}}].

\bibitem{hornreich1975critical}
R.~Hornreich, M.~Luban, and S.~Shtrikman, {\it Critical behavior at the onset
  of k→-space instability on the $\lambda$ line},  {\em Physical Review
  Letters} {\bf 35} (1975), no.~25 1678.

\bibitem{grinstein1981anisotropic}
G.~Grinstein, {\it Anisotropic sine-gordon model and infinite-order phase
  transitions in three dimensions},  {\em Physical Review B} {\bf 23} (1981),
  no.~9 4615.

\bibitem{roberts1998resonant}
J.~Roberts, N.~Claussen, J.~P. Burke~Jr, C.~H. Greene, and C.~Wieman, {\it
  {Resonant magnetic field control of elastic scattering in cold 85Rb}},  {\em
  Physical Review Letters} {\bf 81} (1998), no.~23 5109,
  [\href{http://arxiv.org/abs/physics/9808018}{{\tt physics/9808018}}].

\bibitem{regal2004observation}
C.~Regal, M.~Greiner, and D.~S. Jin, {\it Observation of resonance condensation
  of fermionic atom pairs},  {\em Physical review letters} {\bf 92} (2004),
  no.~4 040403, [\href{http://arxiv.org/abs/cond-mat/0401554}{{\tt
  cond-mat/0401554}}].

\bibitem{nussinov2004bcs}
Z.~Nussinov and S.~Nussinov, {\it {The BCS-BEC crossover in arbitrary
  dimensions}},  \href{http://arxiv.org/abs/cond-mat/0410597}{{\tt
  cond-mat/0410597}}.

\bibitem{Son:2005rv}
D.~T. Son and M.~Wingate, {\it {General coordinate invariance and conformal
  invariance in nonrelativistic physics: Unitary Fermi gas}},  {\em Annals
  Phys.} {\bf 321} (2006) 197--224,
  [\href{http://arxiv.org/abs/cond-mat/0509786}{{\tt cond-mat/0509786}}].

\bibitem{nussinov2006triviality}
Z.~Nussinov and S.~Nussinov, {\it {Triviality of the BCS-BEC crossover in
  extended dimensions: Implications for the ground state energy}},  {\em
  Physical Review A—Atomic, Molecular, and Optical Physics} {\bf 74} (2006),
  no.~5 053622.

\bibitem{Mehen:2007dn}
T.~Mehen, {\it {On non-relativistic conformal field theory and trapped atoms:
  Virial theorems and the state-operator correspondence in three dimensions}},
  {\em Phys. Rev. A} {\bf 78} (2008) 013614,
  [\href{http://arxiv.org/abs/0712.0867}{{\tt arXiv:0712.0867}}].

\bibitem{Nishida:2007pj}
Y.~Nishida and D.~T. Son, {\it {Nonrelativistic conformal field theories}},
  {\em Phys. Rev. D} {\bf 76} (2007) 086004,
  [\href{http://arxiv.org/abs/0706.3746}{{\tt arXiv:0706.3746}}].

\bibitem{chang2007unitary}
S.~Chang and G.~Bertsch, {\it Unitary fermi gas in a harmonic trap},  {\em
  Physical Review A—Atomic, Molecular, and Optical Physics} {\bf 76} (2007),
  no.~2 021603, [\href{http://arxiv.org/abs/physics/0703190}{{\tt
  physics/0703190}}].

\bibitem{von2007bec}
J.~Von~Stecher, C.~H. Greene, and D.~Blume, {\it {BEC-BCS crossover of a
  trapped two-component Fermi gas with unequal masses}},  {\em Physical Review
  A—Atomic, Molecular, and Optical Physics} {\bf 76} (2007), no.~5 053613,
  [\href{http://arxiv.org/abs/0705.0671}{{\tt arXiv:0705.0671}}].

\bibitem{Gegenwart:2008ttt}
P.~Gegenwart, Q.~Si, and F.~Steglich, {\it {Quantum criticality in
  heavy-fermion metals}},  {\em Nature Phys.} {\bf 4} (2008), no.~3 186--197,
  [\href{http://arxiv.org/abs/0712.2045}{{\tt arXiv:0712.2045}}].

\bibitem{Nishida:2010tm}
Y.~Nishida and D.~T. Son, {\it {Unitary Fermi gas, epsilon expansion, and
  nonrelativistic conformal field theories}},  {\em Lect. Notes Phys.} {\bf
  836} (2012) 233--275, [\href{http://arxiv.org/abs/1004.3597}{{\tt
  arXiv:1004.3597}}].

\bibitem{Kaplan:1998tg}
D.~B. Kaplan, M.~J. Savage, and M.~B. Wise, {\it {A New Expansion for
  Nucleon-Nucleon Interactions}},  {\em Phys. Lett. B} {\bf 424} (1998)
  390--396, [\href{http://arxiv.org/abs/nucl-th/9801034}{{\tt
  nucl-th/9801034}}].

\bibitem{Kaplan:1998we}
D.~B. Kaplan, M.~J. Savage, and M.~B. Wise, {\it {Two nucleon systems from
  effective field theory}},  {\em Nucl. Phys. B} {\bf 534} (1998) 329--355,
  [\href{http://arxiv.org/abs/nucl-th/9802075}{{\tt nucl-th/9802075}}].

\bibitem{Mehen:1999nd}
T.~Mehen, I.~W. Stewart, and M.~B. Wise, {\it {Conformal invariance for
  nonrelativistic field theory}},  {\em Phys. Lett. B} {\bf 474} (2000)
  145--152, [\href{http://arxiv.org/abs/hep-th/9910025}{{\tt hep-th/9910025}}].

\bibitem{Kaplan:2009kr}
D.~B. Kaplan, J.-W. Lee, D.~T. Son, and M.~A. Stephanov, {\it {Conformality
  Lost}},  {\em Phys. Rev. D} {\bf 80} (2009) 125005,
  [\href{http://arxiv.org/abs/0905.4752}{{\tt arXiv:0905.4752}}].

\bibitem{Kobach:2018nmt}
A.~Kobach and S.~Pal, {\it {Conformal Structure of the Heavy Particle EFT
  Operator Basis}},  {\em Phys. Lett. B} {\bf 783} (2018) 311--319,
  [\href{http://arxiv.org/abs/1804.01534}{{\tt arXiv:1804.01534}}].

\bibitem{Arav:2019tqm}
I.~Arav, Y.~Oz, and A.~Raviv-Moshe, {\it {Holomorphic Structure and Quantum
  Critical Points in Supersymmetric Lifshitz Field Theories}},  {\em JHEP} {\bf
  11} (2019) 064, [\href{http://arxiv.org/abs/1908.03220}{{\tt
  arXiv:1908.03220}}].

\bibitem{Ardonne:2003wa}
E.~Ardonne, P.~Fendley, and E.~Fradkin, {\it {Topological order and conformal
  quantum critical points}},  {\em Annals Phys.} {\bf 310} (2004) 493--551,
  [\href{http://arxiv.org/abs/cond-mat/0311466}{{\tt cond-mat/0311466}}].

\bibitem{Son:2013rqa}
D.~T. Son, {\it {Newton-Cartan Geometry and the Quantum Hall Effect}},
  \href{http://arxiv.org/abs/1306.0638}{{\tt arXiv:1306.0638}}.

\bibitem{Hoyos:2013qna}
C.~Hoyos, B.~S. Kim, and Y.~Oz, {\it {Lifshitz Field Theories at Non-Zero
  Temperature, Hydrodynamics and Gravity}},  {\em JHEP} {\bf 03} (2014) 029,
  [\href{http://arxiv.org/abs/1309.6794}{{\tt arXiv:1309.6794}}].

\bibitem{Chapman:2015wha}
S.~Chapman, Y.~Oz, and A.~Raviv-Moshe, {\it {On Supersymmetric Lifshitz Field
  Theories}},  {\em JHEP} {\bf 10} (2015) 162,
  [\href{http://arxiv.org/abs/1508.03338}{{\tt arXiv:1508.03338}}].

\bibitem{Arav:2016akx}
I.~Arav, Y.~Oz, and A.~Raviv-Moshe, {\it {Lifshitz Anomalies, Ward Identities
  and Split Dimensional Regularization}},  {\em JHEP} {\bf 03} (2017) 088,
  [\href{http://arxiv.org/abs/1612.03500}{{\tt arXiv:1612.03500}}].

\bibitem{Yan:2022dqk}
Z.~Yan, {\it {Renormalization of supersymmetric Lifshitz sigma models}},  {\em
  JHEP} {\bf 03} (2023) 008, [\href{http://arxiv.org/abs/2210.04950}{{\tt
  arXiv:2210.04950}}].

\bibitem{DearQuantizers}
J.~Kulp, {\it {To appear}},  \href{http://arxiv.org/abs/26xx.xxxxx}{{\tt
  arXiv:26xx.xxxxx}}.

\bibitem{Beg:1984yh}
M.~A.~B. Beg and R.~C. Furlong, {\it {The $\Lambda \phi^4$ Theory in the
  Nonrelativistic Limit}},  {\em Phys. Rev. D} {\bf 31} (1985) 1370.

\bibitem{Jackiw:1991je}
R.~Jackiw, {\it {Delta function potentials in two-dimensional and
  three-dimensional quantum mechanics}},  {\em Diverse topics in theoretical
  and mathematical physics} (1, 1991) 35--53.

\bibitem{Bergman:1991hf}
O.~Bergman, {\it {Nonrelativistic field theoretic scale anomaly}},  {\em Phys.
  Rev. D} {\bf 46} (1992) 5474--5478.

\bibitem{Chapman:2020vtn}
S.~Chapman, L.~Di~Pietro, K.~T. Grosvenor, and Z.~Yan, {\it {Renormalization of
  Galilean Electrodynamics}},  {\em JHEP} {\bf 10} (2020) 195,
  [\href{http://arxiv.org/abs/2007.03033}{{\tt arXiv:2007.03033}}].

\bibitem{Cardy:1992tq}
J.~L. Cardy, {\it {Critical exponents of the chiral Potts model from conformal
  field theory}},  {\em Nucl. Phys. B} {\bf 389} (1993) 577--586,
  [\href{http://arxiv.org/abs/hep-th/9210002}{{\tt hep-th/9210002}}].

\bibitem{Antunes:2026jjf}
A.~Antunes, {\it {Lifshitz critical points meet Zamolodchikov perturbation
  theory}},  \href{http://arxiv.org/abs/2602.12341}{{\tt arXiv:2602.12341}}.

\bibitem{Duval:1994qye}
C.~Duval, P.~A. Horvathy, and L.~Palla, {\it {Conformal Properties of
  Chern-Simons Vortices in External Fields}},  {\em Phys. Rev. D} {\bf 50}
  (1994) 6658--6661, [\href{http://arxiv.org/abs/hep-th/9404047}{{\tt
  hep-th/9404047}}].

\bibitem{Duval:2008jg}
C.~Duval, M.~Hassaine, and P.~A. Horvathy, {\it {The Geometry of Schrodinger
  symmetry in gravity background/non-relativistic CFT}},  {\em Annals Phys.}
  {\bf 324} (2009) 1158--1167, [\href{http://arxiv.org/abs/0809.3128}{{\tt
  arXiv:0809.3128}}].

\bibitem{NullDefects}
R.~S. Erramilli, J.~Kulp, and F.~K. Popov, {\it {Do null defects dream of
  conformal symmetry?}},  \href{http://arxiv.org/abs/2509.04578}{{\tt
  arXiv:2509.04578}}.

\bibitem{nikolic2007renormalization}
P.~Nikoli{\'c} and S.~Sachdev, {\it Renormalization-group fixed points,
  universal phase diagram, and {1/N} expansion for quantum liquids with
  interactions near the unitarity limit},  {\em Physical Review A—Atomic,
  Molecular, and Optical Physics} {\bf 75} (2007), no.~3 033608,
  [\href{http://arxiv.org/abs/cond-mat/0609106}{{\tt cond-mat/0609106}}].

\bibitem{Bekaert:2011qd}
X.~Bekaert, E.~Meunier, and S.~Moroz, {\it {Symmetries and currents of the
  ideal and unitary Fermi gases}},  {\em JHEP} {\bf 02} (2012) 113,
  [\href{http://arxiv.org/abs/1111.3656}{{\tt arXiv:1111.3656}}].

\bibitem{Kravec:2018qnu}
S.~M. Kravec and S.~Pal, {\it {Nonrelativistic Conformal Field Theories in the
  Large Charge Sector}},  {\em JHEP} {\bf 02} (2019) 008,
  [\href{http://arxiv.org/abs/1809.08188}{{\tt arXiv:1809.08188}}].

\bibitem{Favrod:2018xov}
S.~Favrod, D.~Orlando, and S.~Reffert, {\it {The large-charge expansion for
  Schr{\"o}dinger systems}},  {\em JHEP} {\bf 12} (2018) 052,
  [\href{http://arxiv.org/abs/1809.06371}{{\tt arXiv:1809.06371}}].

\bibitem{Kravec:2019djc}
S.~M. Kravec and S.~Pal, {\it {The Spinful Large Charge Sector of
  Non-Relativistic CFTs: From Phonons to Vortex Crystals}},  {\em JHEP} {\bf
  05} (2019) 194, [\href{http://arxiv.org/abs/1904.05462}{{\tt
  arXiv:1904.05462}}].

\bibitem{Hellerman:2023myh}
S.~Hellerman, D.~Krichevskiy, D.~Orlando, V.~Pellizzani, S.~Reffert, and
  I.~Swanson, {\it {The unitary Fermi gas at large charge and large N}},  {\em
  JHEP} {\bf 05} (2024) 323, [\href{http://arxiv.org/abs/2311.14793}{{\tt
  arXiv:2311.14793}}].

\bibitem{Lee:2026azy}
E.~Lee, {\it {Equilibrium Partition Function of Non-Relativistic CFTs in
  Harmonic Trap}},  \href{http://arxiv.org/abs/2603.09856}{{\tt
  arXiv:2603.09856}}.

\bibitem{Beane:2026mxi}
S.~R. Beane, D.~Orlando, and S.~Reffert, {\it {The odd fermion at the edge:
  odd-even staggering in the trapped, unitary Fermi gas}},
  \href{http://arxiv.org/abs/2606.26225}{{\tt arXiv:2606.26225}}.

\bibitem{Son:2008ye}
D.~T. Son, {\it {Toward an AdS/cold atoms correspondence: A Geometric
  realization of the Schrodinger symmetry}},  {\em Phys. Rev. D} {\bf 78}
  (2008) 046003, [\href{http://arxiv.org/abs/0804.3972}{{\tt
  arXiv:0804.3972}}].

\bibitem{Goldberger:2008vg}
W.~D. Goldberger, {\it {AdS/CFT duality for non-relativistic field theory}},
  {\em JHEP} {\bf 03} (2009) 069, [\href{http://arxiv.org/abs/0806.2867}{{\tt
  arXiv:0806.2867}}].

\bibitem{Balasubramanian:2008dm}
K.~Balasubramanian and J.~McGreevy, {\it {Gravity duals for non-relativistic
  CFTs}},  {\em Phys. Rev. Lett.} {\bf 101} (2008) 061601,
  [\href{http://arxiv.org/abs/0804.4053}{{\tt arXiv:0804.4053}}].

\bibitem{Barbon:2008bg}
J.~L.~F. Barbon and C.~A. Fuertes, {\it {On the spectrum of nonrelativistic
  AdS/CFT}},  {\em JHEP} {\bf 09} (2008) 030,
  [\href{http://arxiv.org/abs/0806.3244}{{\tt arXiv:0806.3244}}].

\bibitem{Maldacena:2008wh}
J.~Maldacena, D.~Martelli, and Y.~Tachikawa, {\it {Comments on string theory
  backgrounds with non-relativistic conformal symmetry}},  {\em JHEP} {\bf 10}
  (2008) 072, [\href{http://arxiv.org/abs/0807.1100}{{\tt arXiv:0807.1100}}].

\bibitem{Kachru:2008yh}
S.~Kachru, X.~Liu, and M.~Mulligan, {\it {Gravity duals of Lifshitz-like fixed
  points}},  {\em Phys. Rev. D} {\bf 78} (2008) 106005,
  [\href{http://arxiv.org/abs/0808.1725}{{\tt arXiv:0808.1725}}].

\bibitem{Herzog:2008wg}
C.~P. Herzog, M.~Rangamani, and S.~F. Ross, {\it {Heating up Galilean
  holography}},  {\em JHEP} {\bf 11} (2008) 080,
  [\href{http://arxiv.org/abs/0807.1099}{{\tt arXiv:0807.1099}}].

\bibitem{Gomis:2020fui}
J.~Gomis, Z.~Yan, and M.~Yu, {\it {Nonrelativistic Open String and Yang-Mills
  Theory}},  {\em JHEP} {\bf 03} (2021) 269,
  [\href{http://arxiv.org/abs/2007.01886}{{\tt arXiv:2007.01886}}].

\bibitem{Gomis:2020izd}
J.~Gomis, Z.~Yan, and M.~Yu, {\it {T-Duality in Nonrelativistic Open String
  Theory}},  {\em JHEP} {\bf 02} (2021) 087,
  [\href{http://arxiv.org/abs/2008.05493}{{\tt arXiv:2008.05493}}].

\bibitem{Maxfield:2022hkd}
H.~Maxfield and Z.~Zahraee, {\it {Holographic solar systems and hydrogen atoms:
  non-relativistic physics in AdS and its CFT dual}},  {\em JHEP} {\bf 11}
  (2022) 093, [\href{http://arxiv.org/abs/2207.00606}{{\tt arXiv:2207.00606}}].

\bibitem{Dorey:2022cfn}
N.~Dorey, R.~Mouland, and B.~Zhao, {\it {Black hole entropy from quantum
  mechanics}},  {\em JHEP} {\bf 06} (2023) 166,
  [\href{http://arxiv.org/abs/2207.12477}{{\tt arXiv:2207.12477}}].

\bibitem{Dorey:2023jfw}
N.~Dorey and R.~Mouland, {\it {Conformal quantum mechanics, holomorphic
  factorisation, and ultra-spinning black holes}},  {\em JHEP} {\bf 02} (2024)
  086, [\href{http://arxiv.org/abs/2302.14850}{{\tt arXiv:2302.14850}}].

\bibitem{Mouland:2023gcp}
R.~Mouland, {\it {How to build a black hole out of instantons}},  {\em JHEP}
  {\bf 03} (2024) 002, [\href{http://arxiv.org/abs/2311.13636}{{\tt
  arXiv:2311.13636}}].

\bibitem{Goldberger:2014hca}
W.~D. Goldberger, Z.~U. Khandker, and S.~Prabhu, {\it {OPE convergence in
  non-relativistic conformal field theories}},  {\em JHEP} {\bf 12} (2015) 048,
  [\href{http://arxiv.org/abs/1412.8507}{{\tt arXiv:1412.8507}}].

\bibitem{Simmons-Duffin:2016wlq}
D.~Simmons-Duffin, {\it {The Lightcone Bootstrap and the Spectrum of the 3d
  Ising CFT}},  {\em JHEP} {\bf 03} (2017) 086,
  [\href{http://arxiv.org/abs/1612.08471}{{\tt arXiv:1612.08471}}].

\bibitem{Golkar:2014mwa}
S.~Golkar and D.~T. Son, {\it {Operator Product Expansion and Conservation Laws
  in Non-Relativistic Conformal Field Theories}},  {\em JHEP} {\bf 12} (2014)
  063, [\href{http://arxiv.org/abs/1408.3629}{{\tt arXiv:1408.3629}}].

\bibitem{Dobrev:2013kha}
V.~K. Dobrev, {\it {Non-Relativistic Holography - A Group-Theoretical
  Perspective}},  {\em Int. J. Mod. Phys. A} {\bf 29} (2014) 1430001,
  [\href{http://arxiv.org/abs/1312.0219}{{\tt arXiv:1312.0219}}].

\bibitem{Perroud:1977qh}
M.~Perroud, {\it {Projective Representations of the Schrodinger Group}},  {\em
  Helv. Phys. Acta} {\bf 50} (1977) 233--252.

\bibitem{dobrev1997lowest}
V.~Dobrev, H.-D. Doebner, and C.~Mrugalla, {\it {Lowest weight representations
  of the Schr{\"o}dinger algebra and generalized heat/Schr{\"o}dinger
  equations}},  {\em Reports on mathematical physics} {\bf 39} (1997), no.~2
  201--218.

\bibitem{Pal:2018idc}
S.~Pal, {\it {Unitarity and universality in nonrelativistic conformal field
  theory}},  {\em Phys. Rev. D} {\bf 97} (2018), no.~10 105031,
  [\href{http://arxiv.org/abs/1802.02262}{{\tt arXiv:1802.02262}}].

\bibitem{Fitzpatrick:2011dm}
A.~L. Fitzpatrick and J.~Kaplan, {\it {Unitarity and the Holographic
  S-Matrix}},  {\em JHEP} {\bf 10} (2012) 032,
  [\href{http://arxiv.org/abs/1112.4845}{{\tt arXiv:1112.4845}}].

\bibitem{Penedones:2010ue}
J.~Penedones, {\it {Writing CFT correlation functions as AdS scattering
  amplitudes}},  {\em JHEP} {\bf 03} (2011) 025,
  [\href{http://arxiv.org/abs/1011.1485}{{\tt arXiv:1011.1485}}].

\bibitem{Kulp:2024scx}
J.~Kulp and S.~Pasterski, {\it {Multiparticle states for the flat hologram}},
  {\em JHEP} {\bf 08} (2025) 091, [\href{http://arxiv.org/abs/2501.00462}{{\tt
  arXiv:2501.00462}}].

\bibitem{Chen:2021xkw}
B.~Chen, R.~Liu, and Y.-f. Zheng, {\it {On higher-dimensional Carrollian and
  Galilean conformal field theories}},  {\em SciPost Phys.} {\bf 14} (2023),
  no.~5 088, [\href{http://arxiv.org/abs/2112.10514}{{\tt arXiv:2112.10514}}].

\bibitem{Chen:2023pqf}
B.~Chen, R.~Liu, H.~Sun, and Y.-f. Zheng, {\it {Constructing Carrollian field
  theories from null reduction}},  {\em JHEP} {\bf 11} (2023) 170,
  [\href{http://arxiv.org/abs/2301.06011}{{\tt arXiv:2301.06011}}].

\bibitem{Pal:2016rpz}
S.~Pal and B.~Grinstein, {\it {Weyl Consistency Conditions in Non-Relativistic
  Quantum Field Theory}},  {\em JHEP} {\bf 12} (2016) 012,
  [\href{http://arxiv.org/abs/1605.02748}{{\tt arXiv:1605.02748}}].

\end{thebibliography}\endgroup
\end{document}